\documentclass{JHEP3}
\usepackage{amsmath}

\def\AdSs5{$AdS_5$}
\def\AdS5s5{$AdS_5 \times S^5$}
\def\al{{\alpha^{\prime}}}

\def\NSNS{{$NS\otimes NS$}}
\def\RR{{$R\otimes R$}}

\def\calZ{{\cal Z}}

\def\calN{{\cal N}}
\def\calO{{\cal O}}

\def\calR{{\cal R}}
\def\calD{{\cal D}}

\def\calC{{\cal C}}
\def\Tr{\mbox{Tr}}

\def\hM{{\hat M}}
\def\hN{{\hat N}}

\newcommand{\gym}{g_\text{\tiny YM}}

\newcommand{\ie}{{\it i.e.}}

\newcommand{\be}{\begin{equation}}
\newcommand{\ee}{\end{equation}}
\newcommand{\ba}{\begin{eqnarray}}
\newcommand{\ea}{\end{eqnarray}}
\newcommand{\bea}{\begin{eqnarray}}
\newcommand{\eea}{\end{eqnarray}}
\newcommand{\nn}{\nonumber}

\newcommand{\zero}{\thinspace{}^0\negthinspace}

\DeclareMathOperator{\tr}{tr}
\DeclareMathOperator{\diag}{diag}

\title{D3-branes on the Coulomb branch and instantons}
\author{Michael B. Green and Christian Stahn\\
Department of Applied Mathematics and Theoretical Physics \\
Wilberforce Road, Cambridge CB3 0WA, UK \\
E-mail:
\email{M.B.Green@damtp.cam.ac.uk;}
\email{C.Stahn@damtp.cam.ac.uk}}

\abstract{The relative coefficients of higher derivative
interactions of the IIB effective action of the 
form $C^4$, $(D F_5)^4$, $F_5^8$, $\dots$ (where $C$ is the Weyl 
tensor and $F_5$ is the five-form field strength)
are motivated by supersymmetry arguments.  It is shown that the classical
supergravity solution for $N$ parallel $D3$-branes is
unaltered by this combination of terms.  The
non-vanishing of $\zero C^2$ in this background (where $\zero C$ is the
background value
of the Weyl tensor) leads to effective  $O(\al^{-1})$ interactions, such as
$C^2$ and $\Lambda^8$ (where $\Lambda$ is
the dilatino).  These contain $D$-instanton contributions in
addition to tree and one-loop terms.
The near horizon limit of the $N$ $D3$-brane system
describes a multi-$AdS_5\times S^5$ geometry
that is dual to $\calN=4$ $SU(N)$ Yang--Mills theory spontaneously broken
to $S(U(M_1) \times \cdots \times U(M_r))$.  Here, the
 $D3$-branes are grouped into
$r$ coincident bunches with $M_r$ in each group,
with $\sum_r M_r =N$ and $M_r/N = m_r$ fixed as $N\to \infty$.
The boundary  correlation function of eight $\Lambda$'s is constructed
explicitly.
The second part of the paper considers effects of a
constrained instanton in this large--$N$
Yang--Mills theory  by an extension of the analysis
of Dorey, Hollowood and Khoze of the one-instanton measure at finite $N$.
This makes precise the correspondence with the supergravity
$D$-instanton
measure at leading order in the $1/N$ expansion.  However, the duality
between instanton-induced correlation functions in Yang--Mills theory
and the dual supergravity is somewhat
obscured by complications relating to the structure of
constrained instantons.}

\preprint{DAMTP-2003-57 \\ {\tt hep-th/0308061}}

\begin{document}

\section{Introduction}
\label{intro}

The AdS/CFT correspondence \cite{maldacena,gkp,wittone}
has proved useful in pinning down a number of
features of quantum string theory on the one hand and
supersymmetric Yang--Mills quantum field theory on the other hand.
The literature on this subject is large and it is sometimes
difficult to decipher those aspects of the correspondence that
follow from elementary symmetry principles and those that appear
to contain some more detailed dynamical statements.

An intriguing series of identities that follows from the
correspondence, even though it is not altogether clear why it
should,  concerns properties of Yang--Mills instantons and
$D$-instantons of string theory.  As observed in \cite{bg} and \cite{bgkr}
there
is an obvious qualitative correspondence between these objects,
even in the $SU(2)$ Yang--Mills theory.  Semi-classical expressions for
the
one-instanton contributions to the correlation functions of
operators in the multiplet of superconformal currents in $SU(2)$
superconformal $\calN =4$ Yang--Mills theory  correspond to
the dual supergravity amplitudes.  However, it is only in the
large-$N$ limit of $SU(N)$ Yang--Mills theory that the
correspondence is exact \cite{dhkmv}.  The
measure on the ten-dimensional bosonic instanton moduli space was there
shown to
be that of $AdS_5 \times S^5$, which is identical to the geometric measure
of a point-like $D$-instanton in IIB superstring theory in the presence of
a
large number of coincident $D3$-branes.  Further
evidence for this correspondence is contained in \cite{gk}.
 Half of the sixteen Poincar\'e supersymmetries and sixteen conformal
supersymmetries of $\calN=4$ Yang--Mills theory are broken by the presence
of
an instanton, which therefore carries fermionic moduli corresponding to
eight broken Poincar\'e supersymmetries and eight broken conformal
supersymmetries.  These correspond to the moduli associated with
the super-isometries of the maximally supersymmetric IIB supergravity
background that are broken by the presence of a $D$-instanton.

The superconformal theory is a very special point in the moduli
space of $\calN =4$ theories, with vanishing expectation values
for the scalar fields.  More generally, non-zero expectation
values correspond to displacing the $D3$-branes in transverse
directions.  This breaks the conformal symmetry but the background
still preserves half the Poincar\'e supersymmetries.
The purpose of this paper is to explore properties of
$D$-instantons in this background and to make contact with the
instanton in non-superconformal domains of the $\calN=4$ Yang--Mills
theory.

To begin with, in section~\ref{stringsec}
we will enlarge on the supersymmetry constraints
that determine the leading behaviour of $D$-instantons in the $\al$
expansion
of the IIB effective action. It is known that
$D$-instantons contribute to higher-derivative terms at order
$\al^{-1}$ (where the Einstein--Hilbert term is $O(\al^{-4})$), such as
$C^4$, $C^2\Lambda^8$ and many others (where $C$ is the
ten-dimensional Weyl curvature and $\Lambda$ is the complex dilatino).
Each of these terms is multiplied by a function of the complex scalar
field $\tau$, which is highly constrained by $SL(2,\mathbb{Z})$ invariance.
To understand the effects of these interactions in the multi $D3$-brane
background we need to understand precisely
how the various terms are related by
supersymmetry.  Of particular relevance are those interactions
which depend only on $C$, $\tau$ and
$F_5$ (the five-form field strength) and so
may have  non-zero values in the $D3$ background.
These partners of the familiar  $C^4$ are
interactions of the form
$(DF_5)^4$, $F_5^8$, $C^2\, (DF_5)^2$ among others,
all of which have a
common dependence on $\tau$.  Although we have not completed a full
supersymmetry analysis we  will present compelling evidence that
these terms package together in a simple manner with precisely
defined relative coefficients.   This particular combination of terms
will be shown to leave the background $D3$-brane supergravity
solutions unaltered at $O({\al}^{-1})$.
A simple argument will demonstrate that although $\zero C^4$
is non-zero it cancels with the
background values of the related terms so
that the dilaton one-point function vanishes\footnote{Here and in the following
the background value of any field, $\Phi$, will be denoted by $\zero \Phi$.}.
Furthermore, the graviton and $F_5$ one-point functions will also be
seen to  vanish.
However, $\zero C^2$ is non-vanishing and receives no
corrections from terms involving $\zero F_5$.  Some
relevant properties of the $D3$-brane supergravity
background geometry are described in Appendix~\ref{weylprop} where it is
shown that $\zero C^2 = -4 H^{-3/2}\partial_y^2\partial_y^2 \sqrt H$,
where $H$ is the
standard harmonic function that enters the $D3$-brane solution and is a function of the
six transverse coordinates, $y^i$.
Substituting this expression  into the higher derivative interactions
leads to  $O(\al^{-1})$ interactions of the form $C^2$,
$\Lambda^8$, $G^4$ and many others.  The $D$-instanton contributions to
these
interactions should be dual to corresponding instanton
contributions to correlation functions in large-$N$ $SU(N)$
$\calN=4$ Yang--Mills theory on the boundary.

The $\Lambda^8$  correlation function is used as an illustrative example.
This is obtained from the known effective
$C^2\, \Lambda^8$ supergravity interaction in section~(\ref{lamvertt}) by
attaching
the bulk to boundary dilatino propagator to each of the
eight legs.  This propagator is obtained in terms of the scalar bulk to
boundary propagator by solving the Dirac
equation in the multi $D3$ background.
The same expression is obtained in
section~(\ref{lamzerom}) by directly constructing the profile of
the dilatino in the classical $D$-instanton background.  This is
linear in the Grassmann coordinates corresponding to the
supermoduli for the broken supersymmetries.

The analysis of section~\ref{stringsec} applies to
any geometry resulting from parallel $D3$-branes although the
correspondence with
$\calN=4$ Yang--Mills theory requires  $M_r$
coincident $D3$-branes at transverse
relative positions labelled by $r$, where $M_r/N=m_r$ is
fixed in the limit $N\to \infty$.
These multi-centred backgrounds \cite{klt,kw} include the
special  configurations of $D3$-branes that have been considered in the
literature
\cite{fgpw,bs} which correspond to  continuous distributions of
$D3$-branes,
where the density of the distribution is large in the large-$N$ limit.
These distributions preserve particular subgroups of the
$R$-symmetry group of the superconformal theory.

We will turn in section~\ref{nonconff} to consider the dual $\calN=4$
Yang--Mills theory  in the situation
in which the scalar fields $\varphi^i_{uv}$ have non-zero constant vacuum
expectation values (where $i=1,\dots,6$, and $u,v = 1,\dots, N$ are
colour indices).
The most general configuration is one in which there are $6N$ non-zero
commuting values $\langle\varphi^i_{uv}\rangle=\varphi^i_u \,
\delta_{uv}$.   The instanton moduli space in this situation
was discussed in  \cite{dhk} which focused on the situation with no
degenerate
scalar field vacuum expectation values.   This analysis will be reviewed
in
section~\ref{props}.
Although the quantum theory is still finite,
generic backgrounds of this type are not superconformal
but preserve only the super-Poincar\'e
symmetries. This implies that, in the presence of an instanton,
there are four  exact bosonic moduli $x_0^\mu$ ($\mu=0,1,2,3$),
corresponding to
the broken translation invariance and eight fermionic moduli,
$\eta^A_\alpha$ (where $\alpha =1,2$ is a chiral $SO(3,1)$ spinor
index and $A=1,2,3,4$).
In this situation the  instanton is not an exact solution of
the euclidean theory but only a minimum of the
action  with the non-exact moduli constrained to fixed values.
In other words, there is a non-trivial dependence of the instanton
action on the non-exact moduli, giving rise to a nontrivial
measure.  For small $\gym$ it is possible to study the effects
of such constrained instantons in a systematic manner in perturbation
theory \cite{thooft,affleck,doreyreview}.

In section~\ref{multicent} we will show that the instanton
measure of \cite{dhk} remains valid in cases in which there are
degeneracies of vacuum expectation values. We will reexpress the measure
in terms of an integral representation which is useful for
discussing  symmetry breaking  in which the vacuum
expectation values cluster into $r$ sets of degenerate values with
$M_r$ in each set and with $\sum_r M_r = N$.  In the limit $N\to \infty$
with $M_r/N = m_r$ fixed this configuration should be equivalent to
the near-horizon geometry of a multi-centred configuration of
$D3$-branes, as in \cite{klt,kw}.   The measure will
be expressed as a function of the six non-exact
bosonic moduli, $\chi^i \; (i=1,\dots,6)$ which correspond
to the transverse position of the
$D$-instanton in the dual description
(and reduce to the scale of the instanton and its position on the
five-sphere in the conformal $AdS_5\times S^5$ limit).
We will show that at large $N$ this measure is proportional
to $\partial_\chi^2\partial_\chi^2\, \sqrt H$, where $H$ is the same harmonic
function as the one that enters the dual supergravity background.
In appendix~\ref{schurap}
an alternative discussion is given of the properties of the
measure by  writing it in terms of Schur polynomials.

In addition to the measure, we would like to evaluate
instanton-induced contributions to `minimal' correlation functions
of composite gauge invariant Yang--Mills operators.  These
correlation functions are those in which each operator soaks up at least
one of the eight Poincar\'e supermoduli.  However, there are significant
complications  in this case that do not arise in  the
superconformal case.  These arise from the fact that the
$R$-symmetry group is generically completely broken due to the
vacuum expectation values for the scalar fields.  This leads to
mixing of infinite sets of single-trace operators, as well as
mixing with multi-trace operators. Combined with the fact that
the constrained instantons are not exact solutions this make it
difficult to evaluate the correlation functions.
Further comments on these issues are made at the end of
section~\ref{multicent}
without reaching a firm conclusion.

\section{Type IIB effective action, $D3$-branes and $D$-instantons}
\label{stringsec}

We will first consider the effects of interactions of order $\al^{-1}$
in the presence of $D3$-branes in type IIB superstring theory.  There are
many such
interactions, including the well-known $C^4$ interaction
(where $C$ is the Weyl tensor) which, in string frame, has the form
\cite{gg},
\be
\frac{c_1}{\al}\int d^{10} x \sqrt{\det g}\, e^{-\phi/2}
f^{(0,0)}(\tau,\bar\tau)\,
\left(C^4   + \cdots\right)\, .
\label{rfour}
\ee
Here the specific index contractions
have been suppressed,  $c_1$ is a constant and
$\tau = \tau_1 +i \tau_2 = C^{(0)} + i e^{-\phi}$ is the complex scalar
field (with $C^{(0)}$ being the Ramond--Ramond scalar).  The
fields $\tau$ and $\bar\tau$ parameterise the  coset space
$SL(2,\mathbb{R})/O(2)$.
The
function $f^{(0,0)}(\tau,\bar\tau)$ is defined by the Eisenstein series
\be
f^{(0,0)}(\tau,\bar\tau) = \sum_{(m,n)\ne (0,0)}
\frac{\tau_2^{3/2}} {|m+ n\tau|^{3/2}}\,  ,
\label{maaswave}
\ee
and is invariant under $SL(2,\mathbb{Z})$.  The factor of $e^{-\phi/2}$ in
(\ref{rfour}) is absent in the Einstein frame.
The exact expression was suggested by a variety of arguments
in \cite{gg} and \cite{ggv} and was shown to be a consequence of full
nonlinear supersymmetry in \cite{gs}.   The value of $C^4$ in
(\ref{rfour}) is non-zero in the $D3$-brane background.   The
other interactions indicated by $\cdots$ in
(\ref{rfour}) are those that involve $F_5$ as well as the Weyl curvature.
These are all the interactions that have non-zero values in the
$D3$-brane backgrounds of interest to us here, in which $F_5$
is not constant and the Weyl tensor is
non-vanishing and all other fields are trivial.
We will determine the precise form of these
non-vanishing background terms
in the next sub-section where we will see that they all cancel.

The procedure of  \cite{gs} for determining the fully nonlinear
interactions was to require
closure of the on-shell supersymmetry algebra order by order
in $\alpha'$ which uniquely determines the $O(\al^{-1})$
corrections to the action (up to trivial field redefinitions) as well
as non-trivial and highly nonlinear corrections to the supersymmetry
transformations.
The $\al$ expansion of the IIB effective action has the form
\be
\al^4 S =S^{(0)} + \al^3 S^{(3)} + \cdots\,  ,
\label{actex}
\ee
where $S^{(0)}/\al^4$ is the classical action.
Invariance of the full action
(\ref{actex}) under $\varepsilon^*$ supersymmetry  is ensured only because
the
classical supersymmetry transformations are modified by $O(\al^3)$
corrections so that the complete supersymmetry transformation of any field
$\Phi$
has the form
\be
\delta \Phi =\delta^{(0)}\Phi +\al^3 \delta^{(3)}\Phi + \cdots\, ,
\label{fullsusy}
\ee
where the superscripts indicate the order in the $\al$ expansion.
The expressions
$S^{(3)}$ and $\delta^{(3)}\Phi$ are determined (up to field
redefinitions)
by solving the equation
\be
\delta^{(0)} \, S^{(3)}  + \delta^{(3)} \, S^{(0)}  = 0\, ,
\label{actsusy}
\ee
together with the requirement that the supersymmetry algebra
closes on shell.  This means that
\be
[\delta_1^{(0)}+ \al^3\delta_1^{(3)}, \delta_2^{(0)}+\al^3
\delta_2^{(3)}] \Phi \sim 2{\rm Im}(\varepsilon_1\Gamma^M\varepsilon_2)\,
 D_M\Phi + k(\varepsilon_1,\varepsilon_2)\,\frac{\delta(S^{(0)} + \al^3
 S^{(3)})}{\delta \Phi}
 + O(\al^6)\, ,
\label{closealg}
\ee
so that
the algebra closes on the solutions of the equation of motion defined
by $S^{(0)}+ \al^3 S^{(3)}$.
The quantity $k(\varepsilon_1,\varepsilon_2)$ is bilinear in the
Grassmann parameters  $\varepsilon_1$ and $\varepsilon_2$ which are
sixteen-component $SO(9,1)$ spinors.
The fact that supersymmetry of the effective action
links terms in the $\al$ expansion is no surprise and
is a vital ingredient in the
derivation of the modular forms in the various
interactions.

 Supersymmetry requires the presence of very many
other interactions with classical values that
 vanish in the classical $D3$-brane background.
We will later focus on terms with a factor
of $C^2$, which include (\ref{rfour}) as well as
(again suppressing the index contractions),
\be
\frac{1}{\al}\int d^{10} x \sqrt{\det g}\, e^{-\phi/2}
\left(c_2 f^{(6,-6)}(\tau,\bar\tau)\,
\Lambda^8 C^2\, + c_3 f^{(2,-2)}(\tau,\bar\tau)\, G^4 C^2\, +
\dots\right)\, ,
\label{moreterm}
\ee
where $\Lambda$ is the complex dilatino and $G$ is a complex
combination of the \RR\ and \NSNS\ three-form field strengths.
As with the $C^4$ term, only the traceless part of the curvature (the Weyl
tensor) enters in (\ref{moreterm}).
The  constants $c_1, c_2, \dots$ are easily determined by
linearised supersymmetry combined with modular transformations.
 The modular form
 $f^{(w,-w)}$ transforms under $SL(2,\mathbb{Z})$
with holomorphic weight $w$ and anti-holomorphic weight $-w$, \ie,
\be
f^{(w,-w)}(\tau,\bar\tau) \to \left(\frac{c\tau + d}{c\bar\tau +
d}\right)^\omega \,
f^{(w,-w)}(\tau,\bar\tau) \label{modfor}
\ee
when $\tau \to (a\tau + b)/(c\tau +d)$ (with $ad-bc=1$ and
$a,b,c,d$ integers).
 The modular transformations
of $f^{(w,-w)}$ compensate for the $U(1)$ transformations of the fields
that are induced by $SL(2,\mathbb{Z})$.  We are here fixing the $O(2)\sim
U(1)$
gauge so the scalar fields are restricted to the coset
$SL(2,\mathbb{R})/O(2)$.  In this case the $U(1)$ charges of the fields
are
$q_P= 2$, $q_\Lambda = 3/2$, $q_G = 1$, $q_\psi =1/2$, $q_R =0$ and
$q_{F_5}=0$, where   $P = i \partial\tau/ (2 \tau_2)$ (the scalar field
$\tau$ does not carry a specific $U(1)$ charge).  The total $U(1)$ charge
carried by the fields in any of the terms in (\ref{rfour}) or
(\ref{moreterm}) is $2w$.

Explicitly, $f^{(w-w)}$ is given by  the Eisenstein series,
\bea
\label{eisendef}
f^{(w,-w)}(\tau,\bar\tau) &=&    \frac{1}{ 2^{w}}
\frac{\Gamma\left(w+ \frac{3}{ 2}\right)}{ \Gamma\left(\frac{3}{2}\right)}
\sum_{(m,n) \ne (0,0)}  \frac{\tau_2^{3/2}}{|m+n\tau|^3 }
\left( \frac{m + n \bar\tau}{ m+ n \tau}\right)^w\, .
\eea
This has an expansion  in the string coupling $g =e^{\phi} =
\tau_2^{-1}$ that includes
two power-behaved terms, which correspond to tree-level and
one-loop string theory effects.  In addition there is an infinite
sequence of $D$-instanton and anti $D$-instanton contributions.
The coefficient of each   $D$-instanton  $(K>0$) contribution is of
order $\tau_2^{w}$ while each anti $D$-instanton ($K<0$) contribution
starts with the power $\tau_2^{-w}$.
 In linearised approximation each interaction
term involves a product of $p$ fluctuations of fields,
$\prod_{r=1}^M \Phi_r$, where $p = w + 4$ and the charge $K$ $D$-instanton
contribution is
proportional to
\be
g^{4-p}\, e^{2\pi i K \tau} \, (2\pi|K|)^{1/2}\, \sum_{m|K} \frac{1}{m^2}(
1 +
O(\tau_2^{-1}))\, .
\label{coeffd}
\ee

\subsection{Supersymmetry and the generalised $C^4$ interaction}

In order for the classical multi $D3$-brane background to remain
unaltered by the $O(\al^{-1})$ interactions it is important that the
contributions of these higher derivative interactions
to the one-point functions of the
dilaton, graviton  and $F_5$ all vanish.  For example,
the vanishing of the dilaton one point function requires the terms in the
higher
derivative action (\ref{rfour}) to vanish in the classical background
since they involve the dilaton-dependent
factor $f^{(0,0)}(\tau,\bar\tau)$ which would otherwise alter the
dilaton equation of motion.
Since  the background Weyl tensor,
$\zero C$, the background five-form field strength, $\zero F_5$, and
$\zero D \zero F_5$
are not zero in the non-conformal backgrounds of interest there must be detailed
cancellations between the various terms in
(\ref{rfour}).
 We will now see that
this follows from the BPS condition for the background.

The $32$ components of the type IIB supersymmetry parameters form
$16$-dimensional complex $SO(9,1)$ spinors $\varepsilon^a$
and $\varepsilon^{*\, a}$ ($a=1,2,\dots,16$).
The supersymmetry transformations of the gravitini fields in IIB
supergravity have the form \cite{schwarz}, \cite{hw}
\be
\begin{split}
\delta_\varepsilon \psi_M &= (D_M  + \frac{i}{16\cdot 5!}
\Gamma^{N_1\cdots N_5} F_{N_1\cdots N_5} \Gamma_M)\,\varepsilon + \cdots
  \equiv \calD_M \, \varepsilon  + \cdots \,  , \\
\delta_{\varepsilon^*} \psi^*_M &= (D_M - \frac{i}{16\cdot 5!}
\Gamma^{N_1\cdots N_5} F_{N_1\cdots N_5} \Gamma_M)\,\varepsilon^* + \cdots
  \equiv \calD_M \,\varepsilon^*  + \cdots\,  ,
\end{split}
\label{gravisusy}
\ee
where $M, N, \dotsc = 0, 1, \dots,9$ are ten-dimensional world
indices, $\Gamma^M = e^M_{\ \hM}\, \Gamma^\hM$ and $e_{M\hM}$ is
the ten-dimensional frame field (and $\hM= 0, 1, \dots, 9$ is a
tangent-space
index)\footnote{Tangent
space indices will be labelled by
hatted indices in the following.}.
The $16\times 16$ matrices
$\Gamma^\hM$ are projections of the $SO(9,1)$ acting on chiral spinors
(see appendix~\ref{weylprop} for conventions).
The quantity $\calD$ is defined to
include the
contribution of the five-form field strength.
 The transformations of the dilatini are
\be
\delta_{\varepsilon^*}  \Lambda =
\frac{i}{2}\, \Gamma^M  P_M\, \varepsilon^* +  \cdots  \,  ,\qquad
\delta_{\varepsilon} \Lambda^* =  \frac{i}{2}\, \Gamma^M P^*_M\, \varepsilon
+ \cdots\,.
\label{dilasusy}
\ee
The scalar field   enters into the definition of $P_M =
i \partial_M \tau/ (2 \tau_2)$ and its complex conjugate
 $P^*$.
$D_M$ includes the spin and Christoffel connections, as usual.

The dots in the above equations indicate the contributions of
combinations of
fields that have been suppressed since their precise form  will not be
relevant
for us.  For example, we are ignoring the three-form field
strengths, $G_{MNP}$ and $G^*_{MNP}$,
 as well as various terms quadratic in fermions in the
variation of the gravitini.  Furthermore, the bosonic fields enter
in combination with fermion bilinears that we are suppressing so that they
form
supercovariants (denoted $\hat F_5$, $\hat P$ and ${\hat P}^*$
in \cite{schwarz}).  Such combinations transform under
supersymmetry without derivatives of the parameters
$\varepsilon$ and $\varepsilon^*$.
By acting with $\calD_N$  on equation (\ref{gravisusy}) it is
straightforward to derive the condition
\be
\delta_\varepsilon (\calD_{[M}\psi_{N]} + \cdots) =
(\calR_{MN} + \cdots) \,  \varepsilon\, ,
\label{bpscon}
\ee
(recalling that $\calD$ is defined by (\ref{gravisusy}))
where
\be
\begin{split}
\calR_{MN} = &\frac{1}{8}R_{MNPQ} \Gamma^{PQ} -
\frac{i}{16\cdot 5!}\Gamma^{K_1\cdots K_5} \Gamma_{[M} D_{N]} F_{K_1\cdots K_5}\\
& - \frac{1}{(16 \cdot 5!)^2}
\Gamma^{K_1\dots K_5}\Gamma_{[M} \, \Gamma^{L_1\dots L_5}\Gamma_{N]} \,
F_{K_1\dots K_5}\, F_{L_1\dots L_5} \, .
\label{rmndef}
\end{split}
\ee
 The combination
$(\calD_{[M}\psi_{N]} + \cdots)$ denotes  the supercovariant combination
of ${\calD}_{[M}\psi_{N]}$ and cubic fermionic terms while
$(\calR_{MN} + \cdots)$ denotes
the supercovariant extension of the curvature tensor
(detailed definitions of these
supercovariant contributions  are given in
section 9 of \cite{hw}\footnote{Equation (9.31) in \cite{hw} contain a
 small error in the supercovariant combination that includes
$D_{[M} \psi_{N]}$.}).

The physical content of Type IIB supergravity is contained in a
scalar superfield that is a function of $x^M$, $\theta$ and $\theta^*$,
where the Grassmann coordinates
$\theta$ and $\theta^*$  are sixteen-component $SO(9,1)$
chiral spinors of the same
chirality.  This superfield satisfies an analytic constraint analogous to a
chiral
constraint,
$D^*\Phi =0$, which means that it can be written as a function of
$\theta$ and $\tilde x^M =
x^M - \theta^* \, \Gamma^M \, \theta$ only.
The components in the expansion of $\Phi$ in powers of
$\theta$ are supercovariant combinations of fields and
derivatives of fields, described above.
Symbolically,
\bea
\Phi &=& \tau + \theta\, \Lambda + \theta^2 \, (G+\cdots) + \theta^3\,
(\calD\psi+\cdots) + \theta^4\, (R + DF + FF + \cdots) \nn\\
&& \qquad+ \theta^5
(DD\psi^* + \cdots) + \dots +\theta^8 (DDDD\bar\tau + \cdots) \, ,
\label{phiexpan}
\eea
where dots indicate terms that complete each bracket into
a supercovariant expression.  In this case $\varepsilon$
supersymmetry implies $\delta_\varepsilon \Phi = \varepsilon
\partial \Phi/\partial\theta$.
If the Grassmann parameter is assigned a charge $1/2$ under $U(1)$
transformations all the terms in the superfield (\ref{phiexpan})
have charge $2$ apart from the first ($\theta$-independent) term.
The superfield is a nonlinear generalisation of the scalar superfield of
\cite{hw} and was considered in detail in \cite{hss}.

The detailed form of the term cubic in $\theta$ in (\ref{phiexpan}) is
(dropping the terms indicated by dots)
\be
\theta\, \Gamma^{MNP}\, \theta\,\, \calD_{[M}\psi_{N]} \,\Gamma_P \,
\theta\,
  \,,
\label{cubicr}
\ee
while the term quartic in $\theta$ is proportional to
\be
\theta\, \Gamma^{MNP}\, \theta\, \theta\, \Gamma_P \,
\calR_{MN}\, \theta
= \theta\, \Gamma^{MNP}\, \theta\, \theta\, \Gamma^{QRS}\,
\theta \, \calR_{MNPQRS}\, ,
\label{quarticr}
\ee
where\footnote{
This expression agrees with that deduced from a superfield approach in
version 4 of \cite{hss} (apart from an overall factor of $2$ in the
definition of $\calR$).}
\bea
 &&\calR_{MNPQRS}  =  \frac{1}{8} g_{PS} \, R_{MNQR}+\frac{i}{48}
D_{M}F_{NPQRS}
\nn\\
&&\qquad\quad +\frac{1}{256}\, F_{MNPTU}\, F_{QRS}^{\ \ \ \ \ TU}
 -\frac{1}{768}\, F_{MNSTU}\, F_{QRP}^{\ \ \ \ \ TU}\,
\label{calsix}
\eea
(we are ignoring the three-form field strength and fermion terms).
Equation  (\ref{bpscon}) follows,
after adjusting a relative multiplicative constant,
 by applying $\varepsilon Q$ to
(\ref{quarticr}) and identifying this with  the $\varepsilon$
variation of $\calD_{[M}\psi_{N]}$ in (\ref{cubicr}).
In writing (\ref{calsix}), it is assumed that ${\cal R}_{MNPQRS}$ is
 symmetrised in the manner implied by multiplying by $\theta\, \Gamma^{MNP}\,
 \theta\, \theta\, \Gamma^{QRS}\, \theta$. For example, it should be manifestly
 antisymmetric in $[MNP]$ and $[QRS]$ as well as symmetric under interchange of
 $MNP$ and $QRS$. Fierz rearrangements imply further symmetries, such as the absence
 of double traces. As a result, the components of ${\cal R}_{MNPQRS}$ lie in the sum
 of the ${\bf 1050^+}$ and ${\bf 770}$ representations of $SO(9,1)$\footnote{See
 \cite{pvw} and \cite{pw} for a thorough discussion of group-theoretical aspects of
 $O(\al^{-1})$ terms in the type IIB effective action.}.
The ${\bf 770}$
 is identified with the Weyl tensor, which is the only part of the Riemann tensor
 $R_{MNPQ}$ that survives in (\ref{calsix}), as is well known. The $DF_5$ and $F_5^2$
  terms contribute to the ${\bf 1050^+}$.
The appropriately symmetrised $\zero \calR$ can be written in spinor
basis as
\be
\zero \calR_{[abcd]} \equiv \Gamma^{MNP}_{[ab}\,
\Gamma^{QRS}_{cd]}\,
\zero \calR_{MNPQRS}\, ,
\label{spinr}
\ee
where $a,b,c,d$ are $16$-component chiral $SO(9,1)$
spinor labels.  The antisymmetrisation of these indices follows from
contraction with the Grassmann $\theta$'s in (\ref{quarticr}) and
immediately implies that there are $1820=1050+770$ components.

It is known that in linearised approximation the $O(\al^{-1})$ interactions
in $S^{(3)}$
are contained in the integral of a function of $\Phi(\tilde x,\theta)$
over the sixteen $\theta$'s.  Furthermore, the full nonlinear
supersymmetry uniquely determines the $\tau$-dependent modular
forms $f^{(w,-w)}(\tau,\bar\tau)$.  These statements were combined
in \cite{gs} to deduce the modular function  $f^{(0,0)}$  that multiplies
the
$C^4$ interaction. Together with the analysis of
the preceding paragraph this suggests
that the $C$ and $F_5$ enter into the $O(\al^{-1})$ action
in the combination
\be
S^{(3)}_{\calR^4} = \int d^{10}\tilde x d^{16}\theta \, \det e\,
f^{(0,0)}(\tau,\bar\tau)\, (\theta\, \Gamma^{MNP}\, \theta\,
\theta\, \Gamma^{QRS}\, \theta \, \calR_{MNPQRS})^4 \, .
\label{rfoureps} \ee
This  is  of the form indicated in (\ref{rfour}) but with precisely defined
relative coefficients. The interaction $S^{(3)}_{\calR^4}$ can be expanded as a
combination of $C^4$ and
terms involving $F_5$ and $DF_5$ (other fields being trivial). Since $\calR$
contains a piece of order $F_5^2$ \cite{hss} there  are highly
nonlinear terms, such as $F_5^8$, in (\ref{rfoureps}).   We should
emphasise that we have not proved that (\ref{rfoureps}) contains
all the terms involving only  $C$, $F_5$ and $DF_5$.  That would
require a complete 
analysis of the constraints of nonlinear supersymmetry at
$O(\al^{-1})$, which we have not completed.  However, it is easy
to argue how supersymmetry should determine the form of
(\ref{rfoureps}), including the modular function $f^{(0,0)}$, as follows.
The classical supersymmetries acting on any field
 relate this $\calR^4$ interaction to interactions
of the form $(D\psi+\cdots )\, R^2 \, (D^2\psi^*+ \cdots)$, where
the $\cdots$ inside the brackets again extend these terms to
supercovariant derivatives. In order to determine the modular
function it is necessary to consider the manner in which
supersymmetry mixes  $f^{(0,0)} \calR^4$ with $f^{(1,-1)}\Lambda\, D\psi\,
R^3$.  This requires highly nonlinear modifications of the
supersymmetry transformations, such as the ones deduced in
\cite{gs}.

The superspace analysis of \cite{hss} apparently gives a different
expression from (\ref{rfoureps}), which we will not consider since
it does not reproduce classical tree-level or one-loop string
theory results.

\subsection{Non-renormalization of $D3$ background at $O(\al^{-1})$}

We now want to  understand how the interaction (\ref{rfoureps})
affects BPS solutions of the classical supergravity.  Such
solutions are characterised by the  condition that  the
supersymmetry transformations of the gravitini (\ref{gravisusy})
 vanish (we are only concerned with backgrounds
for which the  three-form field
strengths vanish) so that
\be
D_M \zeta  \equiv (\partial_M + \frac{1}{4} \omega_M^{\hM\hN}\,
\Gamma_{\hM\hN}) \, \zeta
= - \frac{i}{16\cdot 5!} \Gamma^{N_1\cdots N_5} F_{N_1\cdots N_5} \Gamma_M \, \zeta \,
\label{killvect}
\ee
defines a Killing spinor $\zeta$,
with a  conjugate equation for $\zeta^*$.   The vanishing of the
transformations of the
dilatini (\ref{dilasusy}) is automatic if $P_M$ and $P^*_M$ vanish, which
they
do in the $D3$-brane background (but $P_M\ne 0$ in the presence of a
$D$-instanton, which
will be important later). Applying $D_N$ to
(\ref{killvect}) and its conjugate leads to the condition $\calR_{MN} \,
\zeta = 0$
together with its conjugate (setting fermion fields to zero).

Generally, the background will break some of the supersymmetry, even in
the absence of the $D$-instanton.  The $D3$-brane backgrounds of
concern to us break half the supersymmetry.
In order to evaluate the Killing spinors   we shall
decompose the $SO(9,1)$-covariant fields into  $SO(6)\times SO(3,1)$
representations.  The $32 \times 32$ $SO(9,1)$ gamma matrices,
$\hat\Gamma^M$,  are written for $M = i+ 3$ ($i=1,\dots,6$) as
\be
\hat \Gamma^{i+3} = \hat \gamma^i \otimes \gamma_5
                      =  \begin{pmatrix}
                      0 & \Sigma^i\\
                      \bar \Sigma^i & 0
                       \end{pmatrix}
                      \otimes \gamma_5\, ,
\label{gamrep1}
\ee
where $\hat \gamma^i$ are
the $8\times 8$ $SO(6)$ gamma matrices and $\Sigma^i_{AB}$ and
$\bar\Sigma^{i\, AB}$
are $4 \times 4$
matrices.  For $M= \mu =0,1,2,3$,
\be
\hat \Gamma^\mu = I_8  \otimes \gamma^\mu = I_8  \otimes
                                \begin{pmatrix}0 & \sigma^\mu \\
                                \bar \sigma^\mu & 0
                                \end{pmatrix} \, .
\label{gamrep2}
\ee
In our conventions, the ten-dimensional supersymmetry parameter $\varepsilon$
satisfies the chirality constraint $\hat\Gamma_{11}\varepsilon = -\varepsilon$.
It decomposes into $SO(6)\times SO(3,1)$ chiral spinors,
\be
\varepsilon^a = (\varepsilon^A_{+\, \dot\alpha},\, \varepsilon_{-\,A \alpha})\,,
\label{o4o6}
\ee
where $\pm$ indicates the eigenvalue of $\gamma_5$.

The main features of the multi $D3$-brane background are discussed in
appendix~\ref{weylprop} where the well-known solution for
the Weyl tensor and the five-form field
strength  is given in terms of a
harmonic function in the transverse space, $H(y) = 1 + \sum_r M_r
L^4/|y-y_r|^4 \equiv e^{2A}$.
Substituting these into (\ref{killvect}) leads to the solution for
the Killing spinor \cite{duff},
\be
\zeta = H^{-1/8}\, \zeta_+^0 \, ,
\label{epssol1}
\ee
where $\zeta_{+\, \dot\alpha}^{0\, A}$ is an
eight-component constant spinor.
The corresponding condition on $\delta_{\zeta^*} \psi^*_M$
determines the solution  for the Killing spinor
\be
\zeta^* = H^{-1/8}\, \zeta_-^{0\, *}\, .
\label{epssol2}
\ee

In order to determine the $\zero\calR^4$ interactions we will substitute the
background fields into $\calR_{MNRPQS}$ which was defined in
(\ref{calsix}).  First, we note that the $F_5^2$ terms
do not contribute in this background.  The
non-vanishing components of the suitably symmetrised $\calR$ are
\bea
\zero \calR_{\mu\nu i\rho\sigma j} &=& \frac{1}{144}\,
\left( 2 \zero g_{\mu[\rho}  \zero g_{\sigma]\nu} - i
\varepsilon_{\mu\nu\rho\sigma} \right) B_{ij} \,, \nn\\
\zero \calR_{ijklmn} &=& -
\frac{1}{48} \left(6 B_{il} \zero g_{jm} \zero
g_{kn} - i B_i^p \varepsilon_{jklmnp} \right)\,,
\label{compscalr}
\eea
where $B_{ij} \equiv 2D^{(6)}_iD^{(6)}_j A = 2 A_{,ij}-2A_{,i}A_{,j} + \zero g_{ij} A_{,k}
A^{,k}$ is introduced in appendix~\ref{weylprop}.
The terms with $\varepsilon$ tensors in these expressions come from
$\zero D\zero  F_5$ while the remaining terms come from the Weyl tensor,
$C$.
These expressions satisfy duality conditions,
\be
\tfrac{1}{2} \varepsilon_{\mu\nu} {}^{\tau\omega} \zero \calR_{\tau\omega
i\rho\sigma j}
 =  i \zero \calR_{\mu\nu i\rho\sigma j}\, , \qquad
\tfrac{1}{6} \varepsilon_{ijk} {}^{pqr} \zero \calR_{pqrlmn}
 = - i \zero \calR_{ijklmn} \, .
\label{selfduals}
\ee
In the notation of (\ref{spinr}) these
conditions mean that $\zero \calR$
only contains the part that is of definite chirality with respect
to both $SO(3,1)$ and $SO(6)$, namely the part with $\gamma_5 = +1$
and $\hat \gamma_7 =-1$.
This means that in $SO(3,1)\times SO(6)$ spinor notation the  non-zero components
are
\be
\zero \calR_{[(A,\dot\alpha)(B,\dot\beta)(C,\dot\gamma)(D,\dot\delta)]}\, ,
\label{nonzer}
\ee
with no  upper $SO(6)$  or undotted $SO(3,1)$ spinor indices.  As
expected for a 1/2-BPS configuration the effective dimensionality of
each of the four bi-spinor indices is eight.

Now decompose $\theta$ in terms of the $SO(6)\times SO(3,1)$
bi-spinors $\theta_{+\dot\alpha}^A$ and  $\theta_{-\alpha\, A}$.
From (\ref{nonzer}) it follows that the only components of
$\theta$ that contribute to (\ref{quarticr}) are the eight $\theta_+$
components.  Therefore terms in $\theta^{16}\calR^4$
with three or four powers of $\theta_+^4\zero\calR $
vanish  identically since they involve more than eight powers of
$\theta_+$.
This means, for example, that even though $\zero C^4$ is non-zero
the combination of interactions in $\zero\calR^4$ in (\ref{rfoureps})
cancel in the multi $D3$-brane background, which
implies that the dilaton equation of motion is unchanged.  This
cancellation between $\zero C^4$,  $\zero C^2 (\zero D\zero F_5)^2$ and
$(\zero D \zero F_5)^4$ terms can be seen
explicitly by substituting the expressions (\ref{compscalr}).
Similarly,  $\zero\calR^3$ vanishes since it involves twelve powers of
$\theta_+$, which implies that
the graviton and $F_5$ one-point functions vanish.
We conclude that the classical $D3$ background is unaltered by
the $\calR^4$ interactions.
  The first non-vanishing power is $\zero\calR^2$, which
leaves eight powers of $\theta_-$ to be saturated by external
field insertions in scattering amplitudes.  This will be considered in the
next subsection.

\subsection{$O(\al^{-1})$  interactions in $D3$ backgrounds}

We now wish to consider the $D$-instanton contribution to the
$O(\al^{-1})$ interactions in the
$D3$-brane background  and compare them with  the Yang--Mills instanton
contribution (restricted to the one-instanton sector) to the corresponding
correlation functions.  In order to do this we need to substitute
the background values of the fields into the interaction lagrangians
such as , such as $C^4$, $\Lambda^8 C^2$,
$GG^* C^2$, $G^4 C^2$.  To leading order in the
fluctuating fields this leads to interactions
of the form
\be
\int d^4 x \, d^6 y \,  (\det\zero e)\,  \zero C^2\,  {\cal O}\,,
\label{weylform}
\ee
 where ${\cal O}$ is a term of the form
$C^2$, $\Lambda^8$, $GG^*$,  $G^4$,  or one of the many  other
possibilities.
According to the AdS/CFT correspondence
the fields in  ${\cal O}$ couple to the UV Yang--Mills theory on
the boundary at $r\to \infty$.

In writing  (\ref{weylform}) we have used the fact that
\bea
\int d^8\theta_+ \, &&\theta_+\, \Gamma^{M_1N_1R_1}\, \theta_+\,
\, \theta_+\, \Gamma^{M_2N_2R_2}\,
\theta_+ \, \zero \calR_{M_1N_1R_1M_2N_2R_2}\nn\\
&&\theta_+\, \Gamma^{M_3N_3R_3}\, \theta_+\,\, \theta_+\,
\Gamma^{M_4N_4R_4}\,
\theta_+ \, \zero \calR_{M_3N_3R_3M_4N_4R_4}
=c \,  \zero C^2\,
\label{rsqu}
\eea
(where $c$ is a numerical constant)
and gets no contribution from $F_5$ or $D F_5$. To see this we
first note that the Grassmann integrations result in a tensor
proportional to
\be
g^{M_1 M_2}g^{N_1 N_3}g^{R_1 R_4}g^{N_2 R_3}g^{R_2 N_4}
g^{M_3M_4} + {\rm perms}\, .
\label{gras}
\ee
This means that each factor of $\zero \calR$ has an internal pair of
indices
contracted.  However,
any  non-trivial contraction of $\zero \calR$ kills the $\zero D\zero F_5$
term which means that any contraction of the form
$\zero R^M {}_{NPMRS}$  is proportional to the Weyl tensor, $\zero
C_{NPRS}$.  Furthermore, there is a unique non-vanishing contraction of
two Weyl tensors, which is the combination occurring in (\ref{rsqu}).
The factor of $(\det \zero e) \zero C^2$ that
appears in each of the interactions in (\ref{weylform}) is evaluated in
appendix~\ref{weylprop} where it is shown that
\be
(\det \zero e) \zero C^2 = -4 H^{-1} \partial^2 \partial^2 H^{1/2}
\,,
\label{weylres}
\ee
where $H= 1 + \sum_r M_r L^4/|y-y_r|^4$ is the standard  harmonic
function in the transverse space that enters into the $D3$-brane
metric (and $L$ is an arbitrary length scale).

The fields in the composite operators $\calO$ multiply the
remaining Grassmann variables, namely, the eight $\theta_-$
variables.  This determines the tensor structure of these
combinations of fields.  For example, the eight-dilatino term has
the form $\prod_{r=1}^8(\theta_{-\, A_r}^{\alpha_r}
\Lambda_{-\,\alpha_r}^{A_r})$, where $\Lambda_-$ is the
negative chirality component of the complex dilatino.

The preceding analysis applies to any configuration of $N$
parallel $D3$-branes.  However, in order to make contact between
the $\al$ expansion of the classical string theory and
Yang--Mills theory it is necessary to consider $D3$-branes in the
limit in which each brane is close to the horizon of all the other
ones.  After an appropriate rescaling of coordinates the constant
term in $H$  can be dropped in this limit, as usual.

\subsubsection{The dilatino propagator and the $\Lambda^8$ amplitude}
\label{lamvertt}

We will now consider the example of the effective
$f^{(6,-6)}(\tau,\bar\tau)\, \Lambda^8 \, \zero C^2$ interaction in
(\ref{moreterm})
which gives rise to a correlation function of eight $\Lambda_-$ operators on
the $r\to \infty$ boundary.  The tensor structure  in this term is
uniquely specified since the eight components of $\Lambda_-$ are
antisymmetrised.  Explicitly,
 $\int d^8 \theta_-\,
\prod_{r=1}^8\theta_{-\, A_r}^{\alpha_r}\Lambda^{A_r}_{-\,
\alpha_r}\equiv
{(T_8)}_{A_1\cdots A_8}^{\alpha_1\cdots \alpha_8}
\prod_{r=1}^8 \Lambda^{A_r}_{-\, \alpha_r}$, where the tensor
$T_8$ is the unique singlet under $SO(6)$ and $SO(3,1)$ which
 would simply be the $SO(8)$ epsilon tensor,
$\varepsilon^{a_1\cdots a_8}$, if the eight fermionic collective
coordinates were assembled into an $SO(8)$ spinor.

The $D$-instanton contributions are obtained  by considering the Fourier expansion of
$f^{(6,-6)}$. From (\ref{coeffd})  we see that
to leading order in the string coupling constant the $D$-instanton contributions to
$\Lambda_-^8$ are proportional to
\be
f^{(6,-6)}_{K} \sim g^{-6} \, (2\pi |K|)^{1/2}\, \mu(K)\, e^{2\pi iK \tau}
\, .
\label{onedinst}
\ee
The correlation function of eight $\Lambda$'s at points $x_r^{\mu_r}$
on the boundary
$|y|\to \infty$ is obtained by attaching a bulk to boundary propagator to each
$\Lambda_-$ in the interaction vertex. The propagator connecting the
interaction point
 $(x_0^\mu, y_0^i)$ to the appropriate point on the boundary is
obtained by solving the
 ten-dimensional Dirac equation for the dilatino,
\be
\Gamma^MD^{(0)}_M\,  \Lambda=  e^M_\hM \, \Gamma^\hM\,
(\partial^{(0)}_M + \frac{1}{4} \omega_M^{\hM\hN}\,
\Gamma_{\hM\hN})\, \Lambda = -\frac{i}{4\cdot 5!}\, \Gamma^{M_1\cdots M_5}\,
F_{M_1\cdots M_5}\,\Lambda\, ,
\label{dilaeq}
\ee
where the derivatives are with respect to $x_0^\mu$ and $y_0^i$.
Substituting the background fields in the $D$ background results in
the equation
\bea
 \left(\Gamma^i \partial^{(0)}_i +\Gamma^{\mu}
\partial^{(0)}_\mu + \left(\frac{1}{4}+ \frac{1}{2}\gamma^5\right)
\Gamma^i A_{,i}\right)\, \Lambda = 0\,,
\label{tendir}
\eea
where $\gamma^5 \Gamma^i A_{,i}/2$ is the $y_0$-dependent `mass term' while $\Gamma^i
A_{,i}/4$ comes from the spin connections.
Writing $\Lambda= \Lambda_- + \Lambda_+$ gives the coupled equations,
\be
\left(\hat\gamma^i\partial^{(0)}_i - \frac{1}{4} \hat \gamma^i
A_r\right)\Lambda_- = \gamma^\mu \partial^{(0)}_\mu \Lambda_+\,
\label{adsporr}
\ee
and
\be
\left(\hat\gamma^i\partial^{(0)}_i + \frac{3}{4} \hat\gamma^i
A_i\right)\Lambda_+ =
- \gamma^\mu \partial^{(0)}_\mu \Lambda_-\, .
\label{adsmorr}
\ee
It follows that
\be
 \partial_{x_0}^2 \Lambda_- + H^{-1}\left(\hat\gamma^{\hat\imath}
\partial^{(0)}_{\hat\imath} -  \frac{1}{4} \hat\gamma^{\hat\imath}
A_{\hat\imath}\right)\,
\left(\hat\gamma^{\hat\jmath} \partial^{(0)}_{\hat\jmath}  - \frac{1}{4}
\hat\gamma^{\hat\jmath} A_{\hat\jmath}\right)\Lambda_- =0 \,,
\label{adshnew}
\ee
where the hats again indicate flat tangent space indices.
The dilatino solution therefore has the form
\be
\Lambda_{-\, \alpha}^A (x_0,y_0) =\pi^3\, \int d^4 x\,
\hat K_\Lambda (x-x_0;y_0)\, \tilde\Lambda^A_{-\,\alpha}(x)\,,
\label{lamres}
\ee
where  $\tilde\Lambda_-^A(x)$ denotes the value of the dilatino at the point $x$ of the
 boundary at $|y|=r\to \infty$. The dilatino bulk to boundary propagator is given by
\be
\hat K_\Lambda = \hat K(x-x_0;y_0) \, H^{1/8}(y_0)\, ,
\label{ksevdef}
\ee
where $\hat K$ is the  scalar bulk to boundary propagator, which
satisfies the scalar
 Laplace equation,
\be
 (H^{1/2}\partial_{x_0}^2  + H^{-1/2}\partial_{y_0}^2)\, \hat K = 0\,.
\label{hatk}
\ee

The $D$-instanton part of the eight-dilatini amplitude  that emerges
from the $D3$-brane background  is therefore proportional to
\be
\begin{split}
g^2\, e^{-2\pi(iC^{(0)}+ 1/g)}
\int & d^4 x_0\,  d^6 y_0 \, d^8\eta \,
\left(\partial_{y_0}^2\partial_{y_0}^2 H^{1/2}(y_0)\right)\\
& {(T_8)}_{A_1\cdots A_8}^{\alpha_1\cdots \alpha_8}\,
\prod_{r=1}^8\left(\frac{1}{g}\hat K(x_r-x_0;y_0)\,
 \tilde\Lambda^{A_r}_{-\, \alpha_r}(x_r)\right)\, .
\label{lam16res}
\end{split}
\ee
The overall factor of $g^{-6}$ in (\ref{onedinst})  is seen
from (\ref{lam16res}) to come from a factor of $1/g$ for each
external state and a factor of $g^2$ in the measure.
 Since $g$
is identified with $\gym^2/4\pi$ this implies a factor of
 $\gym^4$ in the Yang--Mills instanton measure in accord with
\cite{dhk} and the expression in section~\ref{nonconff}.

\subsubsection{$D$-instanton zero modes}
\label{lamzerom}

The single $D$-instanton induced $\Lambda^8$ correlation function
(\ref{lam16res})
may also be obtained by
a semi-classical analysis of the fermionic zero modes of the
$D$-instanton solution.

A $D$-instanton by itself breaks sixteen of the supersymmetries
\cite{ggp} so that when the fields are set equal
to their background values only
the $\varepsilon$ symmetry is preserved and the $\varepsilon^*$
supersymmetry is broken, resulting in sixteen fermionic
moduli.   This is seen from (\ref{dilasusy}) by recalling that in
a $D$-instanton background $P_M \ne 0$ but $P^*_M=0$ (after continuing to
euclidean signature).
In the $D3$-brane background the
$\varepsilon^*_+$ supersymmetries are already broken by the background
and so they do not correspond to exact $D$-instanton moduli.
Therefore, the net result of adding a $D$-instanton to the $D3$-brane
background is that the eight components  of the $\varepsilon^*_-$
correspond to exact supermoduli since these are exact
symmetries of the background that are broken by the $D$-instanton.

These supermoduli must be soaked up by the operators in
 any $D$-instanton induced correlation function.
 This means that the expectation value of $\calO$
is obtained to leading order in the coupling
by replacing the fields in $\calO$ by their instanton profiles, or zero
modes,
as in the $AdS_5
\times S^5$ case considered in \cite{bgkr}.  The profiles are
obtained    by applying the broken
supersymmetries to the $D$-instanton recursively.
 For example,
identifying $\zeta^*$ with the broken  supersymmetry,
the dilatino zero modes following from  (\ref{dilasusy}) are
\be
\Lambda_{0}  \equiv \delta \Lambda =
 \frac{i}{2}\Gamma^M \, \zero P_M \, \zeta^*_-  \, .
\label{dilvary}
\ee
The quantity  $\zero P_M = e^{-\hat \phi}\partial_M e^{\hat\phi}$
is the classical value of $P_M$ in the $D$-instanton
background and satisfies satisfies $(\zero D^M + 2i \zero Q_M) \zero P_M
=0$
where  $e^{\hat\phi}$ is the
solution of the scalar Green function between two points in the
bulk, $(x,y)$ and $(x_0,y_0)$.  The term $2i \zero Q_M$ arises from
the fact that $P_M$ has $U(1)$ charge $2$ and $\zero Q_M = i\zero
P_M/2$ is  the  composite $U(1)$  gauge potential due to the
$D$-instanton (with
euclidean signature).  For convenience
we normalise the Killing spinor $\zeta^*$ in (\ref{dilvary}) so that
it  has the form
\be
\zeta_-^*  = \left(\frac{H(y_0)}{H(y)}\right)^{\frac{1}{8}}\,
e^{\hat \phi/4}\, \zeta_-^{0\, *} \, ,
\label{killdef}
\ee
where $\zeta_-^{0 \, *}$ is a constant eight-component
Grassmann valued chiral spinor,
which we take to have negative ten-dimensional
chirality in accord with the chirality of the gravitino.
The factor of $e^{\hat\phi/4}$ in the Killing spinor is
due to the $U(1)$ connection in the presence of the $D$-instanton
\cite{bgkr,ggp}\footnote{In \cite{bgkr} and \cite{gg} the factor
of
$e^{\hat\phi/4}$ was incorrectly written as $e^{-\hat\phi/4}$.}.

The expression (\ref{dilvary}) for $\Lambda_0$
is guaranteed to satisfy the
Dirac  equation on the coordinates
$(x,y)$,
\be
\Gamma^M D_M \, \Lambda_0 = \frac{1}{2} \gamma^5\Gamma^iA_i \,
\Lambda_0\, .
\label{dirlamsd}
\ee
The $y$-dependent mass term in this equation has the opposite sign from
the mass term in
(\ref{tendir}) because this is the Dirac equation appropriate to
$\Lambda^*$,
which is conjugate to $\Lambda$.  Furthermore, the $(x_0,y_0)$-dependent
normalisation of the Killing spinor in (\ref{killdef})
has been chosen so  that
$\Lambda_0$ reduces to the suitably rescaled
 bulk to boundary dilatino propagator in the limit $r\to \infty$.   Close
to
the boundary the metric approaches $AdS_5\times S^5$ and $(e^{\hat
\phi} -g)\sim r^{-4} \hat K$, where $\hat K$ is the scalar bulk to
boundary propagator.  Therefore, in this limit $r P_r \to -4
g^{-1}r^{-4} \hat K$.  Together with the fact that $H^{-1/8}(r)
\to r^{1/2}$ and $e^{\hat\phi}\to g$ in this limit it follows from
(\ref{dilvary})
that
\be
\lim_{r\to \infty} \Lambda_0 = r^{-7/2}\, g^{-3/4}\, \hat K(x-x_0,
y_0)\,  H^{1/8}(y_0)\, \zeta^{0\, *}_{-\, A\alpha}\, \tilde
\Lambda^{A\alpha}(x) \equiv  \zeta^{0\, *}_{-\, A\alpha}\,
g^{-3/4} \,
 \Lambda^A_{-\, \alpha}(x_0,y_0) \, ,
\label{dinstm}
\ee
where $\Lambda^A_{-\alpha}$ is the same expression as
(\ref{lamres}).
The eight supermoduli
$\zeta^{0\, *}_-$ are soaked up by the product of eight dilatini in the
correlation function and the
Grassmann integrations generate the tensor $T_8$ contracted into
eight propagators in agreement with  (\ref{lam16res}).  The factor
of $g^{-6}$ in (\ref{lam16res}) is also reproduced.
The factor of $\partial^2\partial^2 H^{1/2}$  in the integrand of
(\ref{lam16res})
should be proportional to the $D$-instanton measure, which we have not
evaluated directly.

\section{Instantons in non-conformal $\calN=4$ supersymmetric Yang--Mills}
\label{nonconff}

We now turn to consider the Yang--Mills dual of the above
superstring description.   The displaced $D3$-branes correspond to
vacuum values for scalar fields that break the gauge symmetry from
$SU(N)$ to $S(U(M_1)\times \dots \times U(M_l))$.  We will again be
interested in the limit $N\to \infty$ with $M_r/N= m_r$ fixed.
Before considering the large-$N$ limit
we will review the analysis of the instanton contribution
by \cite{dhk} of the finite $N$ case with  nondegenerate scalar field
vacuum
expectation values.

\subsection{Review of the one-instanton measure}
\label{props}

With non-zero vacuum expectation values for the
scalar fields the BPST instanton is not a solution of the
euclidean field equations unless its scale is constrained \cite{affleck}.
The constrained instanton
action depends on the scale in a manner that is
controllable in perturbation theory  and gives rise to a nontrivial
measure.  In the case of the $\calN=4$
Yang--Mills theory the measure on the supermoduli space of
such constrained instantons in the presence of non-zero
expectation values for scalar fields is efficiently expressed as a
decoupling limit of the corresponding $D$-brane configuration.  The
background of $N$ parallel $D3$-branes at transverse
positions $\varphi^i_u$ can be obtained by T-duality
on a six-torus
from $N$ $D9$-branes with Wilson lines in the six toroidal directions.
The $D$-instanton arises by T-duality from a $D5$-brane with a
world-volume in the toroidal directions.  The supermoduli of the
instanton are identified with the ground states of the open
strings on the $D5$-brane and the strings joining the $D5$-brane
to the different $D9$-branes.
The open-string ground states on the $D5$-brane (the instanton)
describes a $N=2$
vector supermultiplet consisting of the vector $\chi^i$ and
eight fermionic partners, $\lambda_A^{\dot \alpha}$ and a $N=2$
hypermultiplet that is made up of the
four broken Poincar\'e translations, $x_0^\mu$, and eight
super-translations,
$\eta_\alpha^A$ ($m=1,2,3,4$, $A=1,2,3,4$ and $\alpha =1,2$).
The $D9$-$D5$ open-string ground states are the
moduli $w_{u\, \dot \alpha}$, $\bar w^{u\, \dot \alpha}$
and their super-partners $\mu_{Au}$, $\bar \mu_A^u$ which fit into a $N=1$
hypermultiplets of the $D5$-brane (the $D5$-$D9$ system being $1/4$ BPS).

The partition function defined by these variables is
given by
\be
\calZ = \pi^{-6} \gym^4 \int d^8 \eta^A_\alpha\, d^4 x_0^\mu\,
\hat\calZ\, ,
\label{resdor}
\ee
where $\hat{\cal Z}$ is the centred partition function,
\be
\begin{split}
\hat\calZ = & 2^{-2N-1} \pi^{-6N-9} \, \int d^6\chi\, d^{2N}w\, d^{2N}\bar
w \,
d^8\lambda\, d^{4N}\mu\, d^{4N}\bar\mu\, d^3D \exp \Bigl[
\bar w^{\dot \alpha}  \tilde\chi^2 w_{\dot \alpha}
\\
& - iD^c(\tau^c)^{\dot \alpha} {}_{\dot \beta} \bar w^{\dot \beta}  w_{\dot \alpha}
- 2 g_0^{-2}D^2+
2\sqrt2i\pi\bar\mu^{A} \tilde\chi_{AB}
\mu^{B}+i\pi(\bar\mu^{A}  w_{\dot \alpha}
+ \bar w^{\dot \alpha} \mu^A)\lambda^{\dot \alpha}_A \Bigr] \,
\end{split}
\label{modint}
\ee
and $D^c$ ($c=1,2,3$) is a standard auxiliary coordinate.
The vacuum expectation values
$\varphi^i_u$ are contained in the combination
\be
\tilde\chi^i_u = \chi^i - \varphi^i_u \,.
\label{chitildef}
\ee
The dimensional coupling constant in (\ref{modint}) is defined by
$g_0 = \gym^2 \, \al^{-2}$.  We are interested in the
decoupling limit $\al\to 0$ in which
system reduces to the field theoretic Yang--Mills instantons and so we
will
set  $g_0\to \infty$ from here on.   In the superconformal
theory the six bosonic moduli $\chi^i$ describe the unit vector on the
five-sphere and  the
instanton scale size.
It is now convenient, following \cite{dhk},
 to integrate the moduli that carry a gauge
index.
The $\mu^A_u$ and $\bar\mu^{A\, u}$ integrals can be evaluated
by completing the square of the fermionic terms, leading to
\be
\begin{split}
\hat\calZ = & 2^{-2N-1} \pi^{-2N-9} \int d^6\chi\, d^{2N}w\, d^{2N}\bar w \,
d^8\lambda\,  d^3D \, \Bigl( \prod_u \tilde\chi_u^4 \Bigr) \\
&\times \exp \left[-\tfrac{i\pi}{2\sqrt2}\lambda_{\dot \alpha\, A}\bar
w^{\dot \alpha}
(\tilde\chi^{-1})^{AB}w_{\dot \beta}\lambda^{\dot \beta}_B -\bar
w^{\dot \alpha}  \tilde\chi^2 w_{\dot \alpha}-
iD^c(\tau^c)^{\dot \alpha}_{\ \dot \beta}
\bar w^{\dot \beta}  w_{\dot \alpha}
-2g_0^{-2}D^2 \right] \,,
\label{midint}
\end{split}
\ee
The $w^{\dot \alpha}_u$ and $\bar w^{\dot \alpha\, u}$ integrations
can now be performed, giving
\be
\hat\calZ =  \frac{1}{2\pi^9} \int d^6\chi\, d^3D\, d^8\lambda\, \prod_{u=1}^N
\frac{\tilde\chi_u^4}{\tilde\chi_u^4+(D+\Xi_u)^2}\, ,
\label{semfin}
\ee
where the fermion bilinear $\Xi$ is defined by
\be
\Xi^c_u= \frac{\pi}{2\sqrt2}(\tau^c)^{\dot \alpha}_{\ \dot \beta}
\lambda_{\dot \alpha\, A}(\tilde\chi_u^{-1})^{AB}\lambda^{\dot \beta}_B\ .
\label{xidef}
\ee
It should be noted that if all the vacuum values are equal so that
$\tilde \chi^i_u = \chi^i$ and $\Xi_u = \Xi$, the fermionic terms in
the integrand of
(\ref{xidef}) can be eliminated by shifting the variable $D^c$.
In that case there is superconformal invariance and the $\lambda$ integration causes
the measure to vanish. It is useful to change variables from  $\lambda_{A\, \dot\alpha}$
to $\bar\xi^A_{\dot\alpha}$ defined by $\lambda_{A\, \dot\alpha} =
 \chi_{AB}\,\bar\xi^B_{\dot \alpha}$.  The measure then reduces to the usual
 superconformal measure $d^8\mu\, d^8 \bar\xi\, d^4 x\, d\rho/\rho^{-5}$, where
 $\rho =  |\chi|^{-1}$.

The eight fermionic integrations pick out the term quartic in $\Xi$ in the
Taylor
expansion of the integrand of (\ref{semfin}) in powers of $\Xi$.
This term has a factor of $D^{-4}$ which makes the
$d^3D$ integration convergent and,
as shown in \cite{dhk}, the result of performing these integrals
is that the measure can be written as an integral over the six
components of $\chi^i$ in the form
\be
\hat\calZ = -\frac{1}{16 \pi^3}
\int d^6 \chi\, \partial_\chi^2\, \partial_\chi^2
\, I_N(\chi)\, ,
\label{fincalz}
\ee
where $\partial^2$ is the flat
Laplace
operator in the six-dimensional $\chi^i$ space.
The function $I_N$ depends on $\chi^i$
via its dependence on the $N$ quantities,
\be
\label{idef}
x_u = \tilde\chi^2 \equiv (\chi - \varphi_u)^i\, (\chi - \varphi_u)^i \,,
\ee
and is given by
\be
I_N (\chi) \equiv I_N(x_1,\dots,x_N) =
\sum_{u=1 \dots N} x_u^{-1}
\prod_{\substack{v= 1\dots N \\ v \neq u}}
\frac{x_v^2}{x_v^2-x_u^2}\, .
\label{indef}
\ee

The integral in (\ref{fincalz}) was evaluated in \cite{dhk} using Gauss'
law, giving
\be
\hat\calZ = \hat\calZ_f + \hat\calZ_\infty\,,
\label{divcalz}
\ee
where $\hat\calZ_f$ comes  from surface integrals around the points $x_u = 0$
 ($\chi = \varphi_u$).
  In the non-degenerate case considered in
\cite{dhk} the strength of each of these singular contributions is
one so that $\hat\calZ_f = N$. The contribution $\hat\calZ_\infty$
comes from  the surface at $|\chi| \to \infty$, which is the
small-instanton limit and
corresponds to a $D$-instanton at the boundary of $AdS_5$.  The asymptotic behaviour
$I_N \sim k_N |\chi|^{-2}$ as $|\chi|\to \infty$ leads to
$\hat\calZ_\infty = -k_N$, so that
\be
\hat\calZ= N - k_N\, ,
\label{fulres}
\ee
 where
\be
k_N = \frac{2\,\Gamma(N+1/2)}{\Gamma(N)\, \Gamma(1/2)}\, .
\label{kndef}
\ee
For future reference we note that as $N\to \infty$,
\be
k_N = \frac{2}{\sqrt{\pi}}N^{1/2}
\left(1-\frac{1}{8N} + O\left(\frac{1}{N^2}\right)\right)
\, ,
\label{stirlingap}
\ee
as follows from Stirling's approximation.
Some properties of $I_N$ are elucidated by expressing it in terms
of Schur functions as discussed in appendix~\ref{schurap}.

\subsection{Multi-centred configurations}
\label{multicent}

We now want to extend the analysis of \cite{dhk} to configurations
for which there is a superstring dual that can be studied in
semi-classical approximation.  This first requires us to
demonstrate that the expression for the
instanton measure applies to the situation in which the eigenvalues
are degenerate.  We will denote by  $\varphi^i_r$ an eigenvalue that
has degeneracy $M_r$ so that $\sum_{r=1}^l M_r = N$.
In this situation, the centred partition function (\ref{semfin}) becomes
\be
\hat\calZ= \frac{1}{2\pi^9} \int d^6 \chi \; d^3 D \; d^8 \lambda \;
F \,, \qquad
F = \prod_{r=1}^l \left[\frac{x_r^2}{x_r^2+(D+\Xi_r)^2}
\right]^{M_r}\, .
\label{fderf}
\ee
Instead of repeating similar steps to those of \cite{dhk} it is
useful to introduce the integral representation,
\be
F = \prod_{r=1}^l \frac{x_r^{2M_r}}{\Gamma(M_r)}
\int_0^\infty dy_r \; y_r^{M_r-1}
\exp\left[-y_r(x_r^2+(D+\Xi_r)^2) \right]\, .
\label{intone}
\ee
In this representation it is straightforward to perform the $D^c$
integrations,
\be
\begin{split}
\int d^3 D \; F =
\pi^{3/2}
\int_0^\infty &\left( \prod_{r=1}^l \frac{x_r^{2M_r}
y_r^{M_r-1}}{\Gamma(M_r)} dy_r \right)
\Bigl(\sum_r y_r\Bigr)^{-3/2}\\
&\times\exp\left[\frac{(\sum_r y_r \Xi_r)^2}{\sum_r y_r}-
\sum_r y_r(x_r^2+\Xi_r^2) \right]
\label{intdrep}
\end{split}
\ee
Performing the integrations over the eight factors of $\lambda$ picks out
the
term quartic in  $\Xi$, which has the form
\be
\begin{split}
\int d^8 \lambda \; d^3 D \; F=
\frac{\pi^{3/2}}{8}
\int_0^\infty &
\left(\prod_r \frac{x_u^{2M_r}y_r^{M_r-1}}{\Gamma(M_r)} dy_r\right)
\Bigl(\sum_r y_r\Bigr)^{-7/2}
\exp\Bigl[-\sum_r y_rx_r^2\Bigr]\\
&\times\sum_{rstu} y_ry_sy_ty_u
\int d^8\lambda \; (\Xi_r-\Xi_s)^2(\Xi_t-\Xi_u)^2\\
=
\frac{\pi^{3/2}}{8}
\int_0^\infty &
\left( \prod_r \frac{z_r^{M_r-1}}{\Gamma(M_r)} dz_r \right)
\Bigl(\sum_r \frac{z_r}{x_r^2}\Bigr)^{-7/2}
\exp\Bigl[-\sum_r z_r\Bigr]\\
&\times\sum_{rstu} \frac{z_r z_s z_t z_u}{x_r^2 x_s^2 x_t^2 x_u^2}
\int d^8\lambda \; (\Xi_r-\Xi_s)^2(\Xi_t-\Xi_u)^2 \,,
\end{split}
\label{fourxiag}
\ee
where $z_r=y_r x_r^2$. The Grassmann integrations over the eight components of $\lambda$
can be performed making use of the identity
\be
\int d^8\lambda \, \prod_{i=1}^4 \lambda\, \tau^{c_i}\, \Sigma^{i_i}\,
\lambda
=
\begin{aligned}[t]
2^8\, \delta^{c_1c_2} & \delta^{c_3c_4}
\left(\delta^{i_1i_2}\delta^{i_3i_4}  - \delta^{i_1i_4}\delta^{i_2i_3} -
\delta^{i_1i_3}\delta^{i_2i_4}\right)\\
&+ \text{\rm permutations of 1234} \,
\end{aligned}
\label{lambdint}
\ee
(where the normalisation $\int d\theta\, \theta=2$ for Grassmann
integration has been adopted).
This can be used to show that
\begin{multline}
\sum_{rstu}  \frac{z_r z_s z_t z_u}{x_r^2 x_s^2 x_t^2 x_u^2}
\int d^8\lambda \; (\Xi_r-\Xi_s)^2(\Xi_t-\Xi_u)^2
= 3 \cdot 2^{13} \pi^4 \sum_{rstu}
\frac{z_rz_sz_tz_u}{x_r^3x_s^3x_t^3x_u^3}\\
\Bigl(
-x_rx_s + 2 x_r \tilde\chi_s\!\cdot\!\tilde\chi_t
+5\tilde\chi_r\!\cdot\!\tilde\chi_s \tilde\chi_t\!\cdot\!\tilde\chi_u
+6\frac{x_tx_u}{x_rx_s}(\tilde\chi_r\!\cdot\!\tilde\chi_s)^2
-12\frac{x_u}{x_r}
\tilde\chi_r\!\cdot\!\tilde\chi_s \tilde\chi_r\!\cdot\!\tilde\chi_t
\Bigr)\\
=- 2^9 \pi^4 \left(\sum_r \frac{z_r}{x_r^2}\right)^{7/2}
\partial_\chi^2\partial_\chi^2 \left(\sum_r \frac{z_r}{x_r^2} \right)^{1/2}\,.
\label{toshow}
\end{multline}
Hence
\be
\int d^8 \lambda \; d^3 D \; F
= 2^9 \pi^8 \,\partial_\chi^2\partial_\chi^2\, I_{\{M_r\}} \,,
\label{finint}
\ee
where
\be
I_{\{M_r\}}(x_1,\dots,x_l) = -\frac{1}{8\pi^{7/2}} \int_0^\infty \left(
\prod_r \frac{z_r^{M_r-1}}{\Gamma(M_r)} dz_r
\right)
\Bigl(\sum_r \frac{z_r}{x_r^2} \Bigr)^{1/2}
\exp\Bigl[-\sum_r z_r\Bigr]\, .
\label{schewin}
\ee
This expression can also be obtained from
\be
\begin{split}
I_{\{M_r\}}(x_1,\dots,x_l) &= \frac{1}{2^5\pi^5} \int \frac{d^3 D}{D^4} \;
\prod_{r=1}^l \left(\frac{x_r^2}{x_r^2+D^2} \right)^{M_r}\\
&= \frac{1}{8\pi^4} \int_0^\infty \frac{dD}{D^2} \;
\prod_{r=1}^l \int_0^\infty dz_r \frac{z_r^{M_r-1}}{\Gamma(M_r)}
\exp\left[-z_r\left(1+\frac{D^2}{x_r^2}\right)\right]  \,
\end{split}
\label{schwinger}
\ee
by evaluating the gaussian integral over $D$ (dropping an irrelevant
$\chi$-independent divergence at $D=0$). In the non-degenerate case
($M_r=1$) the first line of (\ref{schwinger})
coincides with an  expression in  \cite{dhk} where it was shown to
be equal to $I_N$  (\ref{indef}), \ie, $I_{\{M_r=1\}} \equiv I_N$.
Conversely, $I_{\{M_r\}}$ coincides with the limit
\be
I_{\{M_r\}}(x_1,\dots, x_l) =
I_N(\underbrace{x_1,\dots,x_1}_{M_1}, \underbrace{x_2,\dots,x_2}_{M_2}, \;\; \dots \;\;, \underbrace{x_l,\dots,x_l}_{M_l}) \,.
\label{convers}
\ee
The integral representation of $I_{\{M_r\}}$ in (\ref{schewin})
is useful for estimating its behaviour in various limits.

\paragraph{Large-distance limit $|\chi| \equiv r \to \infty$.}
Consider an arbitrary configuration with $l=N$ centres, \ie, all $M_r=1$ (which includes
degenerate cases). Writing
\be
x_u = (\chi-\varphi_u)^2 = r^2(1+\varepsilon_u) \,, \qquad
\varepsilon_u = -2 \frac{\chi\cdot\varphi_u}{r^2} +
\frac{\varphi_u^2}{r^2} \,,
\label{larger}
\ee
we have
\be
I_N(x_1,\dots,x_N) = -\frac{1}{8\pi^{7/2}} \frac{1}{r^2}
\int_0^\infty dz_1\cdots dz_N \;
\left(\sum_{u=1}^N \frac{z_u}{(1+\varepsilon_u)^2} \right)^{1/2}
\exp\Bigl[-\sum_u z_u\Bigr]\, .
\label{newress}
\ee
The limit $r\equiv|\chi| \gg \varphi_u$ corresponds to $\varepsilon_u\to 0$. To leading
order in the $\varepsilon_u$, the integrand is a function of $s=z_1+\cdots +z_N$ only,
and we obtain
\bea
I_N(x_1,\dots,x_N) &=& -\frac{1}{8\pi^{7/2}} \frac{1}{r^2}
\int_0^\infty ds \; s^{1/2} e^{-s}
\int_{0\le z_1+\cdots+z_{N-1} \le s} dz_1 \cdots dz_{N-1}
+O(\varepsilon_u)
\nn\\
&=& \left(-\frac{1}{16\pi^3}\right) \frac{2}{\sqrt{\pi}}
\frac{\Gamma(N+1/2)}{\Gamma(N)} \frac{1}{r^2} +
O(\varepsilon_u)\, ,
\label{zeroeps}
\eea
which is the same result as in \cite{dhk},
\be
I_N(x_1,\dots,x_N) =
\left(-\frac{1}{16\pi^3}\right) \frac{k_N}{r^2}
+O(\varepsilon_u) \, .
\label{inlarger}
\ee

\paragraph{Behaviour of $I_{\{M_r\}}$ close to any of the $l$ centres.}
We can choose, without loss of generality,  $x_1=r^2 \ll x_2, \dots, x_l$. In that case
\be
\begin{split}
I_{\{M_r\}} (x_1,&\dots,x_l) = -\frac{1}{8\pi^{7/2}} \frac{1}{r^2}
\int_0^\infty dz_1 \frac{z_1^{M_1-\tfrac{1}{2}}}{\Gamma(M_1)}
\exp(-z_1) \\
&\times \int_0^\infty dz_2\cdots dz_l \;
\frac{z_2^{M_2-1}\cdots z_l^{M_l-1}}{\Gamma(M_2)\cdots\Gamma(M_l)}
\left(1+\sum_{r=2}^l \frac{z_r}{z_1}\frac{r^4}{x_r^2} \right)^{1/2}
\exp\Bigl[-\sum_{r=2}^l z_u\Bigr] \,.
\label{closecent}
\end{split}
\ee
To zeroth order in $r^2/x_r$, the integrals over $z_2, \dots, z_l$
give unity, and
\bea
I_{\{M_r\}}(x_1,\dots,x_l) &=&
\left(-\frac{1}{16\pi^3}\right) \frac{2}{\sqrt{\pi}}
\frac{\Gamma(M_1+1/2)}{\Gamma(M_1)} \frac{1}{r^2} + O(r^4/x_r^2)
\label{inclose1}
\\
&=&
\left(-\frac{1}{16\pi^3}\right) \frac{k_{M_1}}{r^2}
+ O(r^4/x_r^2)
\label{inclose2}
\eea
Using this as an estimate for $I$ close to each of the centres allows the explicit
computation of the centred partition function (\ref{fincalz}),
\bea
\hat\calZ = \hat\calZ_f+\hat\calZ_\infty =
\sum_{r=1}^l k_{M_r} - k_N \,,
\label{newasym}
\eea
where the last term comes from the integral around the point at $r=\infty$.  This is a
small extension of the result of \cite{dhk}.

\paragraph{Large-$N$ limit with fixed $M_r/N$.}
We will now discuss the large-$N$ limit of $I$ and
$\hat \calZ$ in order to see how these results fit in with
the known behaviour of $D$-instanton
effects in type IIB supergravity.
Consider multi-centred configurations with large number of VEVs at
each centre such that $M_r = m_r N$ with all $m_r>0$
fixed and satisfying $\sum_r m_r=1$.
This leads to an integral that can be evaluated using the saddle-point
method,
\begin{multline}
I_{\{M_r\}}(x_1,\dots,x_l) = \\
-\frac{1}{8\pi^{7/2}}
\int_0^\infty \left(
\prod_r \frac{dz_r}{\Gamma(M_r)}\right)
\Bigl(\sum_r \frac{z_r}{x_r^2} \Bigr)^{1/2}
\exp\left[\sum_r \left(-z_r +(M_r-1) \log z_r\right)\right] \,.
\label{isaddle}
\end{multline}
The exponent becomes stationary at $z_r=M_r-1$. Changing variables
to $\hat z_r = z_r-M_r-1$ gives
\be
\begin{split}
I_N = -\frac{1}{8\pi^{7/2}} \prod_r \int_{1-M_r}^\infty
& \frac{d\hat z_r}{\Gamma(M_r)}
\left(\sum_r \frac{M_r-1+\hat z_r}{x_r^2} \right)^{1/2}
\\
& \times (M_r-1)^{M_r-1} e^{1-M_r}
\exp\left[-\sum_r \frac{\hat z_r^2}{2(M_r-1)}\right]
\label{inlargen1}
\end{split}
\ee
\be
\xrightarrow{N\to\infty}
\left(-\frac{1}{16 \pi^3}\right) \frac{2}{\sqrt{\pi}}
\left(\sum_r \frac{M_r}{x_r^2} \right)^{1/2}
= -\frac{1}{8 \pi^3\sqrt{\pi}}\,
\left(\sum_r \frac{M_r}{|\chi -\varphi_r|^4} \right)^{1/2}\, ,
\label{inlargen2}
\ee
where Stirling's approximation has been used.
We therefore see that $I_N$ is proportional to $\sqrt{H}$, where
$H$ is a harmonic function in the six-dimensional space spanned by
$\chi^i$.   It is of course no accident that this is the harmonic
function that enters into the metric for $N$ $D3$-branes
(which is reviewed in appendix~\ref{weylprop}) in the near-horizon
large-$N$ limit.

The scale size of the instanton is defined by $\rho^2 = \bar
w^{u\dot \alpha}w_{u\dot \alpha}$.  Evaluating its expectation
value with the measure defined by (\ref{modint})
results in $\rho^2 = \langle \bar w w \rangle = 2\sum_r M_r/x_r$.
This reduces to $\rho^2 = 2N/r^2$ in the $|r|\to\infty$ (small instanton)
limit and approaches $\infty$ in the various infra-red
limits at $\tilde \chi_r \sim 0$ where the instanton should reduce to a
BPST instanton of the $SU(M_r)$ subgroup.  This behaviour is in qualitative
agreement with expectations.

When the degeneracy of eigenvalues is finite at a number of sites the large-$N$
 limit of $I_N$ is not equal to $\sqrt H$. However, as shown at the end of
 appendix~\ref{schurap}, the asymptotic behaviour of $I_N$ as a function of $r$
 matches that of $\sqrt H$ at least up to terms of order $r^{-9}$.
 The derivation
 relies on writing $I_N$ in terms of Schur polynomials although this result
 can presumably also be extracted from the integral representation
 (\ref{indef}).

\subsection*{Comments on correlation functions in $\calN=4$  Yang--Mills theory}

The matching of correlation functions in the Yang--Mills theory
with the supergravity amplitudes is not so straightforward.  Even in
the absence of the instanton the $D3$ background metric is
complicated to describe in terms of the boundary Yang--Mills
theory.  As described in \cite{klt} and \cite{kw} the classical
scalar field expectation
values generically break the $SO(6)$ $R$-symmetry, leading
to an infinite tower of non-vanishing single-trace chiral
primary operators, $\calO_l^{(i_1\dots i_l)} = \Tr(\varphi^{(i_1}\dots
\varphi^{i_l)})$ as well as muti-trace products of these operators,
 which mix with each other.  The tensor indices in these expressions
are defined by the values of the scalar field expectation values.
These operators couple to the
trace of $h_{MN}$ on the
five-sphere,  where  $h_{MN} = g_{MN} - g_{AdS}$ is the
 deviation of the metric from the
$AdS_5\times S^5$ metric,  $g_{AdS}$.  The multi-trace components are
essential for generating  terms nonlinear in $M_r$ in the expansion of
$H^{1/2}$ in powers of $1/r$.

In addition,  even in the superconformal theory, where the scalar
fields $\varphi^i$ have
zero classical expectation values,
the presence of an instanton leads to expectation
values which are
proportional to  products of two fermion
moduli, $\varphi^{[AB]}_{uv} \sim \bar \mu^{A}_u\mu^{B}_v$
(where $u$ and $v$ are $SU(N)$ indices taking
$N$ values).
Substituting in the expression for the chiral
primary operators and
integrating over these fermions gives the multi-trace
condensate corresponding to the string-frame metric for an instanton in
$AdS_5\times S^5$.  Similar condensates arise for superconformal
chiral  descendents
for which the relevant single trace operators are given by expressions
such as
 $\calC_l^{(i_1\dots i_l)} = \Tr\left((\zero F^-)^2\varphi^{(i_1}\dots
\varphi^{i_l)}\right)$,  $\hat\Lambda_l^{A i_1\dots i_l } =
\Tr(\sigma^{\mu\nu}\zero F^- \lambda^A \varphi^{i_1}\dots
\varphi^{i_l})$, etc. (where the appropriate symmetrisations of
indices  is assumed).   In the superconformal theory $l$ would indicate
the Kaluza--Klein mode  on the five-sphere to which the operator
couples.

When $\varphi$ has both a classical expectation value and  and an
instanton-induced fermion bilinear the situation is even more
complicated.  This  has not been analysed in detail but certain
qualitative features are apparent.  The
combination of operators that couple to the dilaton $\zero\calC$ in this
background is a sum of
single-trace operators $\calC_l$ multiplied by factors of $\calO_{l_r}$.
In order to match the supergravity expression for the $D$-instanton
solution of the dilaton this combination has
to be equal to $\hat K$, the solution of
the ten-dimensional Laplace equation (\ref{hatk}).  Similarly, the
dilatino couples to $\zero\hat \Lambda$ which is a sum of
single-trace operators $\hat \Lambda_l$
multiplied by factors of $\calO_{l_r}$.  As usual, the instanton
profile of $\hat\Lambda_l$ is linear in the fermionic
collective coordinates for the broken supersymmetries,
$\eta^A_\alpha$.  This should lead to a non-zero correlation
function of eight $\hat \Lambda$'s that matches the eight-dilatini correlation
function (\ref{lam16res}).

\section{Summary}
\label{discuss}

In this paper we considered aspects of higher derivative
interactions of type IIB superstring theory in the background of
a collection of parallel $D3$-branes.
To begin with we considered the higher derivative interactions of the
IIB effective  action at
$O(\al^{-1})$ that are functions only of $C$, $\tau$ and $F_5$ (as
well as the metric) and which might
therefore be non-zero in the $D3$ background.  An argument that combined
supersymmetry and $SL(2,\mathbb{Z})$ invariance
 was used to package all these terms
into a highly nonlinear expression of the form $\al^{-1}\int d^{10}x\, \det e\,
f^{(0,0)}\, \calR^4$.  A full proof that there are no additional terms
involving only these fields 
has not been completed.  We saw that $\calR$ possesses an elegant
self-duality property in the $D3$ background from which it follows that
$(\zero\calR)^4=(\zero\calR)^3 =0$ and so the background is not
affected by the order $\al^{-1}$ interactions.  More precisely, the one-point
functions of the dilaton, graviton and five-form field strength all
vanish, which is in accord with stringy  intuition.
The non-zero value of the curvature leads to a
non-zero value of $\sqrt{\zero g}\zero \calR^2\sim \sqrt{\zero g}\zero C^2$
proportional to $H^{-1}\partial_y^2 \partial_y^2 H^{1/2}$, where $H$
is the harmonic function that enters in the classical background solution.
There are no terms quadratic in the five-form
background field or its derivative.  As a result, there are terms in the
effective action of the form $C^2$, $\Lambda^8$, $G^4$ and many
others, which all have known $D$-instanton contributions.  The
$D$-instanton contribution to the correlation function of eight
$\Lambda$'s on the $|y|\to \infty$ boundary of the near-horizon geometry
was explicitly
determined.  This involved constructing the bulk to boundary
propagator for $\Lambda_-$, which is given in terms of the product of
a Killing spinor and the scalar bulk to boundary propagator, $\hat K$.
This is also the
structure obtained from the  fermionic zero modes in the $D$-instanton
background.
Special solutions of the scalar Laplace equation have
been discussed in the literature for continuous distributions of
$D3$-branes that have some residual symmetry (see, for example,
\cite{klt,fgpw,bs}).

The second part of the paper (section~\ref{nonconff})
 considered the effects of an instanton
in non-conformal regions of the moduli space of $\calN=4$
supersymmetric $SU(N)$ Yang--Mills theory at large $N$.  We
argued that the expression for the measure of the
constrained one instanton moduli space of \cite{dhk}
applies to degenerate  cases in which eigenvalues of the scalar fields
coincide.  In the limit $N\to \infty$  with $M_r/N$ fixed (where $M_r$
is the degeneracy of eigenvalues with value $\varphi_r$) the
measure on the six scalar moduli, $\chi^i$, was found to be proportional
to $\partial_\chi^2\partial_\chi^2 H^{1/2}$.   This
is the same factor as appeared in the $D$-instanton measure
(with $y^i$ identified with $\chi^i$, apart from a dimensional
constant).

The comparison of instanton induced correlation functions of gauge invariant
operators with corresponding supergravity amplitudes is more
problematical.  The background geometry is described by a
complicated sum of multi-trace operators of the boundary theory,
involving the classical vacuum values of the scalar fields.
The presence of the instanton induces additional expectation
values of the Yang--Mills scalar fields that are bilinear in the
infinite number of non-exact fermionic moduli.  We have not sorted
out the full effect of these vacuum values but expect that the
 correlation functions should match those of the supergravity.

The effective IIB supergravity action contains a great deal
of information concerning multiply charged $D$-instanton
contributions. This should provide information about multi  Yang--Mills
instantons
which we have not considered explicitly in these backgrounds.    The
agreement between the two sides indicates, for example, that the
Yang--Mills measure should contain a factor of $\sum_{m|K}1/m^2$,
just as in the $AdS_5\times S^5$ case.

\acknowledgments

We are very grateful to Gary Gibbons, Robert Helling, Stefano Kovacs, Hugh Osborn,
 Malcolm Perry and Paul Townsend for useful discussions.
CS would like to thank the Studienstiftung des deutschen Volkes for financial support.

\appendix

\section{Properties of the $D3$-brane background supergravity solution}
\label{weylprop}

We will here review some useful properties of the supergravity
background considered in the main text and also define notation and
conventions.

Type IIB supergravity admits multi-centre $D3$-brane solutions
\cite{duff} with a metric of the form
\bea
ds^2 = \zero g_{MN} dx^M dx^N =  H^{-1/2} \eta_{\mu\nu} dx^\mu dx^\nu
+ H^{1/2} \delta_{ij} dy^i dy^j \, ,
\label{metric10d}
\eea
where $\eta_{\mu\nu} = \diag(-1,+1,+1,+1)$ and the
indices take values $M=0,\dots,9$, $\mu=0,\dots,3$, $i=1,\dots,6$.
The self-dual five-form field strength is given by
\begin{equation}
\begin{split}
\zero F_5 &= (1+*) dc^{(4)} \,, \\
c^{(4)} &= H^{-1} dx^0 \wedge dx^1 \wedge dx^2 \wedge dx^3 \,,
\end{split}
\label{f5bg}
\end{equation}
the dilaton is constant, and all other fields are set to zero. The field
equations
\bea
\zero R^{MN} - \tfrac{1}{2} \zero g^{MN} \zero R &=& \frac{1}{16\cdot 6}
\zero F^{M M_2\cdots M_5} \zero F^N {}_{M_2\cdots M_5}
\label{einstein}
\eea
and $d \zero F_5 = 0$ are satisfied if $H$ is a harmonic function of
the transverse coordinates $y^i$,
\begin{equation}
H^{1/2}\, D^M D_M H(y) =  \delta^{ij} \partial_i \partial_j H(y) = 4\pi^3
L^4 \sum_r M_r \delta^{(6)} (y^i - y^i_r)
\label{hdefeq}
\end{equation}
where $L$ is a length scale and $M_r$ are integers.
The source terms represent a superposition of parallel $D3$-branes,
with $M_r$ coincident branes at $y=y_r$. The solution
\begin{equation}
H = 1 + \sum_r \frac{L^4 M_r}{|y-y_r|^4}\equiv e^{2A} \,,
\label{harmdef}
\end{equation}
where $|y-y_r|^2 = (y^i - y_r^i)^2$,
leads to a metric which is asymptotically flat.
The harmonic condition implies
$A_{,i} {}^i = -2 A_{,i} A^{,i} + (\text{$\delta$ function terms})$.

The non-zero components of the Christoffel connection for the
metric (\ref{metric10d}) are
\begin{equation}
\zero\Gamma^\mu_{\nu i} = - \tfrac{1}{2} \delta^\mu_\nu A_{,i} \, , \;\;
 \zero\Gamma^i_{\mu\nu} = \tfrac{1}{2} \zero g_{\mu\nu} A^{,i} \, , \;\;
 \zero\Gamma^i_{jk} = \tfrac{1}{2} \bigl( \delta^i_j A_{,k} +
 \delta^i_k A_{,j} - \zero g_{jk} A^{,i} \bigr)\, ,
\label{chriss}
\end{equation}
and the Riemann tensor has non-zero components
\begin{equation}
\begin{split}
\zero R_{\mu\nu\rho\sigma} &= -\tfrac{1}{2} A_{,i} A^{,i} \zero
g_{\mu[\rho} \zero g_{\sigma]\nu} \, , \\
\zero R_{\mu i\nu j} &= \tfrac{1}{4} \zero g_{\mu\nu} (-3 A_{,i}
A_{,j} + 2 A_{,ij} + \zero g_{ij} A_{,k} A^{,k} ) \, , \\
\zero R_{ijkl} &= \left( \zero g_{i[l} A_{,k]j} - \tfrac{1}{2}
\zero g_{i[l} A_{,k]} A_{,j} + \tfrac{1}{4} \zero g_{i[l} \zero g_{k]j}
A_{,m} A^{,m} \right) - (i \leftrightarrow j) \, ,
\end{split}
\end{equation}
where antisymmetrisations are with unit weight.
The nonvanishing components of the Ricci tensor are
\begin{equation}
\zero R_{\mu\nu} = - \zero g_{\mu\nu} A_{,i} A^{,i} \,,\qquad
\zero R_{ij} = \zero g_{ij} A_{,k} A^{,k} - 2 A_{,i} A_{,j} \,,
\label{riccidef}
\end{equation}
and the the curvature scalar vanishes,
\be
\zero R =0\, .
\label{vanr}
\ee
 The Weyl tensor has components
\begin{equation}
\label{weylbg}
\begin{split}
\zero C_{\mu i\nu j} &= \tfrac{1}{4} \zero g_{\mu\nu} B_{ij} \, ,\\
\zero C_{ijkl} &= \tfrac{1}{2} \left( -\zero g_{i[k} B_{l]j} + \zero
g_{j[k} B_{l]i} \right) \,.
\end{split}
\end{equation}
Here we have introduced the symmetric traceless tensor
\begin{equation}
B_{ij}\equiv 2D^{(6)}_iD^{(6)}_jA = 2 A_{,ij}-2A_{,i}A_{,j} + \zero g_{ij} A_{,k} A^{,k}
\,,
\label{bijdef}
\end{equation}
where $D^{(6)}_i$ signifies the covariant derivative with respect to the
six-dimensional transverse space.
The only quadratic diffeomorphism invariant that can be constructed
from the Weyl tensor is
\bea
\zero C_{MNPQ} \zero C^{MNPQ} = 2 \tr B^2 = -4 H^{-3/2} \partial^2
\partial^2 H^{1/2} \,.
\eea

The field strength $F_5$ in (\ref{f5bg}) is self-dual by
construction. We follow the convention where the Hodge dual is defined
by
\bea
(* F)_{M_1\cdots M_5} &=& \frac{1}{5!} \varepsilon_{N_1\cdots N_5 M_1
\cdots M_5} F^{N_1\cdots N_5} \,,
\eea
and $\varepsilon_{M_1\cdots M_{10}}$ are the components of the volume
form, \ie, $\varepsilon_{0\cdots 9} = H^{1/2}$. The non-vanishing
components of $\zero F_5$  are
\begin{equation}
\label{f5bgcomps}
\zero F_{\mu\nu\rho\sigma i}  = -2
  \varepsilon_{\mu\nu\rho\sigma}A_{,i} \,, \qquad
  \zero F_{jklmn} = 2 \varepsilon_{ijklmn} A^{,i} \,,
\end{equation}
where $\varepsilon_{\mu\nu\rho\sigma}$ and $\varepsilon_{ijklmn}$
denote the components of the four- and six-dimensional volume forms,
$\varepsilon_{0123} = H^{-1}$ and $\varepsilon_{123456} =
H^{3/2}$. The covariant derivative of $F_5$ has components
\begin{equation}
\label{df5bg}
\begin{split}
\zero D_i \zero F_{j\mu\nu\rho\sigma} =&  - \varepsilon_{\mu\nu\rho\sigma}
B_{ij} \\
\zero D_i \zero  F_{jklmn} =&  (2 A_{,i} {}^p - A_{,i} A^{,p})
\varepsilon_{pjklmn}
+ 5 A^{,p} A_{[,j} \varepsilon_{klmn]ip} \, .
\end{split}
\end{equation}

We introduce an orthonormal frame
\begin{equation}
e_\mu^{\hat\mu} = H^{-1/4} \delta_{\mu}^{\hat\mu} \,, \qquad
e_i^{\hat\imath} = H^{1/4} \delta_i^{\hat\imath}
\label{vielbein}
\end{equation}
where hats denote flat indices.
The connection one-form then has components
\begin{equation}
\omega_\mu^{\hat\nu\hat\jmath} = -\tfrac{1}{2} \delta_{\mu}^{\hat\nu}
\delta_j^{\hat\jmath} A^{,j} \,, \qquad
\omega_i^{\hat\jmath\hat k} =
 \tfrac{1}{2} (\delta_i^{\hat\jmath} \delta^{\hat k l}
 - \delta_i^{\hat k} \delta^{\hat\jmath l}) A_{,l} \, .
\label{spinconn}
\end{equation}
The $32 \times 32$ Dirac matrices $\hat\Gamma_{\hat M}$ satisfy
\begin{equation}
\{ \hat\Gamma_{\hat M} , \hat\Gamma_{\hat N} \} = 2 \eta_{\hat M \hat N}
\, ,
\qquad \eta = \diag(-1,+1,\dots,+1) \, .
\label{clifford}
\end{equation}
The chirality of a ten-dimensional spinor is defined by the
eigenvalue of $\hat\Gamma_{11}= \hat\Gamma^{\hat 0} \cdots \hat\Gamma^{\hat 9}$,
which has $\hat\Gamma_{11}^2 = 1$.
The matrices $\hat\Gamma$ can be expressed as
\be
\hat \Gamma_{\hat M} = \begin{pmatrix}
                    0 & \Gamma_{\hat M} \\
                     \bar\Gamma_{\hat M} & 0
                       \end{pmatrix}\, ,
\label{chiralgamm}
\ee
where $\Gamma_{\hat M\, ab}$ and $\bar\Gamma_{\hat M}^{ab}$
are $16 \times 16$ matrices which
act on chiral spinors (with upper and lower indices corresponding to $\pm$
chiralities).  A Gamma matrix with a curved index is obtained using the
frame field,
 $\Gamma_M = e_M^{\hat M} \Gamma_{\hat M}$.  In these conventions,
expressions such as
$\lambda_1^*\, \Gamma^{M_1\cdots M_r} \lambda_2$ transform as $SO(9,1)$
tensors.  If the chiralities of $\lambda_1$ and $\lambda_2$ are equal $r$
must be odd while if the chiralities are unequal $r$ must be
even.  The quantity $\Gamma^{M_1\cdots M_r}$ is defined so that when
when the $M_i$ are all distinct, it is  equal to $\Gamma^{M_1}
\bar \Gamma^{M_2}\cdots \Gamma^{M_r}$ if $r$ is odd and $\Gamma^{M_1}
\bar \Gamma^{M_2}\cdots \bar\Gamma^{M_r}$ if $r$ is even (with a
corresponding definition of $\bar\Gamma^{M_1\cdots M_r}$).  For
convenience the bars are omitted from the $\Gamma$'s in the text
since their positions  are always obvious by context.

The covariant derivative acting on a spinor
$\varepsilon$ is
\begin{equation}
D_\mu \varepsilon = \partial_\mu \varepsilon - \tfrac{1}{4} A^{,j}
\Gamma_{\mu j} \varepsilon \, , \qquad
D_i \varepsilon = \partial_i \varepsilon + \tfrac{1}{4} A^{,j}
\Gamma_{ij} \varepsilon \,.
\label{dspinor}
\end{equation}
The equations of motion of type IIB supergravity are invariant under
32 supersymmetries which form two Majorana-Weyl spinors
$\varepsilon_1$, $\varepsilon_2$ of the same ten-dimensional
chirality. The chirality is linked to the choice of sign in $F_5 = \pm
* F_5$, and in our conventions $\Gamma_{11} \varepsilon_{1,2} = -
\varepsilon_{1,2}$. We use the complex combinations $\varepsilon =
\varepsilon_1 + i \varepsilon_2$, $\varepsilon^* = \varepsilon_1 - i
\varepsilon_2$. In the $D3$ background the
supersymmetry variation of the gravitino is
\bea
\delta_\varepsilon \psi_M = D_M \varepsilon + \frac{i}{16 \cdot 5!} \Gamma^{N_1
\cdots N_5} \zero F_{N_1 \cdots N_5} \Gamma_M \varepsilon = D_M
\varepsilon - \tfrac{1}{4} \Gamma^i A_{,i}\, \Gamma^{(5)} \,
 \Gamma_M \varepsilon \,,
\label{deltapsi}
\eea
with a corresponding equation for the variation of $\psi_M^*$
(with $\Gamma^{(5)} = i \Gamma^{\hat 0\hat 1\hat 2 \hat 3} =
\gamma^5\otimes 1$).
The background preserves those supersymmetries with $\delta_\zeta
\psi_M = 0$ and $\delta_{\zeta^*}
\psi_M^* = 0$. In terms of the
projected Killing spinors $\zeta_\pm = \tfrac{1}{2}
(1\pm \Gamma^{(5)}) \zeta$, this is equivalent to the conditions
\begin{equation}
\begin{aligned}
\partial_\mu \zeta_+ &= 0 \,, \\
\partial_\mu \zeta_- &= \tfrac{1}{2} \Gamma^j A_{,j} \Gamma_\mu
\zeta_+ \,,
\end{aligned}
\qquad\qquad
\begin{aligned}
\left(\partial_i + \tfrac{1}{2} A^{,j} \Gamma_{ij} -
\tfrac{1}{4}A_{,i} \right) \zeta_+ &= 0 \,,\\
\left(\partial_i + \tfrac{1}{4} A_{,i} \right) \zeta_- &= 0 \,.
\end{aligned}
\label{ksecomps}
\end{equation}
For generic distributions of parallel $D3$-branes
the solution of these conditions is given by the 16
Killing spinors
\bea
\zeta_- &=& 0 \,, \qquad \zeta_+ = H^{-1/8} \zeta^0_+\nn\\
\zeta^*_+ &=& 0 \,, \qquad \zeta^*_- = H^{-1/8} \zeta^{0*}_- \,,
\label{susybg}
\eea
where $\zeta^0_+$ and $\zeta^{0*}_-$ are constant eight-component spinors.

\section{Instanton measure and Schur polynomials}
\label{schurap}

In the following, we will present an alternative discussion of some of the properties of
the measure of section \ref{multicent}, starting from
\bea
\hat\calZ &=& -\frac{1}{16\pi^3} \int d^6 \chi \partial_\chi^2
 \partial_\chi^2 \, I_N \,,
\nn\\
I_N(x_1,\dots,x_N) &=&
\sum_{u=1 \dots N} x_u^{-1}
\prod_{\substack{v = 1\dots N \\ v \neq u}}
\frac{x_v^2}{x_v^2-x_u^2} \,.
\label{indefap}
\eea
In order to consider degenerate configurations of vacuum expectation
values,
we will need to study the function $I_N(x_1,\dots,x_N)$ with its arguments
set
equal in clusters. Similarly, a large-distance asymptotic expansion is
equivalent to
taking all arguments of $I_N$ close to one. Both limits can be analysed by
rewriting
$I_N$ in terms of Schur polynomials, which avoids the apparent
singularities at $x_u=x_v$.

\subsection{Properties of the function $I_N$}
\label{inprops}

It will prove useful to consider generalisations of
$I_N$ defined by
\ba
I_N^{(d)}(x_1,\dots,x_N)  &=&
\sum_{u=1 \dots N} x_u^d
\prod_{\substack{v = 1\dots N \\ v \neq u}}
\frac{x_v^2}{x_v^2-x_u^2} \,,
\ea
where $d$ is an arbitrary integer. The $I_N^{(d)}$ are symmetric
homogenous rational functions.
A series of manipulations shows that they have particularly
simple representations in terms of Schur polynomials,
\begin{equation}
\begin{split}
I_N^{(d)} &=
\frac{1}{\prod_{w<t}(x_w^2-x_t^2)}
\sum_u x_u^d
\prod_{v\neq u}x_v^2
\frac{\prod_{w<t}(x_w^2-x_t^2)}{
\prod_{v \neq u}
(x_v^2-x_u^2)}\\
&= \frac{1}{\prod_{w<t}(x_w+x_t)}
\frac{1}{\Delta}
\sum_u
(-1)^{N-u} x_u^d
\prod_{v \neq u} x_v^2
\prod_{\substack{w<t \\w,t \neq u}}
(x_w^2-x_t^2)
\end{split}
\end{equation}
with $\Delta = \prod_{w<t}(x_w-x_t)$. Noting that
$\prod_{\substack{w<t \\w,t \neq u}}(x_w^2-x_t^2)$ is the Vandermonde
determinant of the $x_v^2, \; v \neq u$, it follows that
\bea
I_N^{(d)} &=&
\frac{1}{\prod_{w<t}(x_w+x_t)}
\frac{1}{\Delta}
\det
\left(
\begin{array}{cccc}
x_1^{2(N-1)} & x_2^{2(N-1)} & \cdots & x_N^{2(N-1)} \\
x_1^{2(N-2)} & x_2^{2(N-2)} & \cdots & x_N^{2(N-2)} \\
\vdots & \vdots & \ddots & \vdots \\
x_1^{4} & x_2^{4} & \cdots & x_N^{4} \\
x_1^{2} & x_2^{2} & \cdots & x_N^{2} \\
x_1^d & x_2^d & \cdots & x_N^d
\end{array}
\right) \,.
\label{inddet}
\eea
Now recall that for any $N$-tuple
$\lambda=(\lambda_1,\dots,\lambda_N)$, the quotient
\bea
S_\lambda (x_1,\dots,x_N) &=& \frac{1}{\Delta}
\det \left[ (x_v^{\lambda_u+N-u})_{u,v=1\dots N}\right]
\eea
is a polynomial which in the case of a partition
(\ie where $\lambda_u\ge\lambda_{u+1}$) is called the Schur polynomial
associated with $\lambda$ \cite{macd}.
For $d=2,4,\dots,2N-2$, two rows of the determinant in (\ref{inddet})
coincide, so $I_N^{(d)}=0$. In all other cases, after taking out a factor
of $(\prod_u x_u)^{\min(2,d)}$ from the determinant and reordering the
rows, one obtains a Schur polynomial.
For example, for $d<0$, the result is
\bea
I_N^{(d)}=\frac{(\prod_u x_u^{d})S_\lambda}{\prod_{u<v}(x_u+x_v)}
\label{indneg}
\eea
with $\lambda = (N-1-d,N-2-d,\dots,1-d,0)$.
Among other things, this implies that $I_N^{(d)}$ is not, as it may
appear from the definition, singular at $x_u=x_v$.

An important property of $I_N^{(d)}$ that we shall need later is its
value when all arguments are set equal to unity. The Schur polynomials
are the characters of the irreducible representations of
$SU(N)$, and their values at $x=1$ are given by the dimension formula
\bea
S_\lambda (1,1,\dots,1) &=&\prod_{\substack{u,v = 1\dots N \\ u<v}}
\frac{\lambda_u-\lambda_v+v-u}{v-u} \,.
\eea
For example, for $d=-1$,
\bea
S_{(N,N-1,\dots,3,2,0)}
&=& 2^{(N-1)(N-2)/2}
\frac{(2N-1)!!}{(N-1)!}\,,
\eea
and therefore
\be
I_N^{(-1)}(1,1,\dots,1) = 2^{1-N}\frac{(2N-1)!!}{(N-1)!} \equiv k_N \,.
\ee
It is straightforward to compute similarly that for all integer $d$,
\bea
I_N^{(d)}(1,1, \dots,1)&=&\binom{N-\tfrac{d}{2}-1}{-\tfrac{d}{2}}
\equiv
\frac{(1-\tfrac{d}{2})\cdots((N-1)-\tfrac{d}{2})}{(N-1)!}.
\label{ind1}
\eea
For later reference note that the large-$N$ limit of this expression
(obtained by using Stirling's approximation) has the form
\be
I_N^{(d)}(1,1, \dots,1) =
\frac{\Gamma(N-d/2)}{\Gamma(1-d/2)\Gamma(N)} =
\frac{N^{-d/2}}{\Gamma(1-d/2)}
\left(1+O\left(\frac{1}{N}\right)\right) \,.
\label{ind1asym}
\ee

\subsection{Large-$N$ limit}

We will first consider the simplest non-trivial vacuum configuration,
namely, one in which $M_1$ of the vacuum expectation values of the
scalar fields take one value $\varphi_u = a_1$ (suppressing the
six-dimensional vector index) and the remaining $N-M_1 =M_2$ ones
take a second value $\varphi_u = a_2$, with $M_1 a_1+M_2 a_2=0$.
This corresponds to setting $x_1=x_2=\dots=x_{M_1}=y\equiv (\chi-a_1)^2$
and $x_{M_1+1}=\dots=x_N=z\equiv (\chi-a_2)^2$. Without loss of
generality,
we take the instanton modulus $\chi$ to be in a region with $y<z$. As a
first
step, we consider $I_N(y_1,\dots,y_{M_1},z_{1}, \dots, z_{M_2})$ with
non-degenerate
arguments satisfying $y_u<z_w$ and expand
\be
\label{expas}
\begin{split}
I_N(y_1,&\dots,y_{M_1},z_{1}, \dots, z_{M_2})\\
=&\sum_{u=1 \dots M_1} \frac{1}{y_u}
\prod_{\substack{v = 1\dots M_1 \\ v \neq u}}
\frac{y_v^2}{y_v^2-y_u^2}
\prod_{w = 1\dots M_2}
\left(1-\frac{y_u^2}{z_w^2}\right)^{-1}\\
&+\sum_{u=1 \dots M_2} \frac{1}{z_u}
\prod_{w=1\dots M_1}
\left(1-\left(1-\frac{y_w^2}{z_u^2}\right)^{-1} \right)
\prod_{\substack{v = 1\dots M_2 \\ v \neq u}}
\frac{z_v^2}{z_v^2-z_u^2}\,.
\end{split}
\ee
The products over $w$ can be rewritten using the complete symmetric
functions $h_r$,
\be
\label{hrdef}
h_r(x_1,\dots,x_n) = \sum_{i_1+\cdots+i_n=r}
x_1^{i_1}\cdots x_n^{i_n} \,, \qquad
\prod_{i=1}^n (1-x_i t)^{-1} =
\sum_{r\ge 0} h_r(x_1,\dots,x_n) t^r \, ,
\ee
which leads to
\be
\label{taylexpand}
\begin{split}
I_N(y_1,&\dots,y_{M_1},z_{1}, \dots, z_{M_2})\\
=& \sum_{r=0}^\infty
I_{M_1}^{(2r-1)}(y_1,\dots,y_{M_1}) \;
h_r(z_1^{-2},\dots,z_{M_2}^{-2})\\
&+ (-)^{M_1} \sum_{r=0}^\infty y_1^2 \cdots y_{M_1}^2 \;
h_r(y_1^2,\dots,y_{M_1}^2)\;
I_{M_2}^{(-2(M_2+r)-1)}(z_1,\dots,z_{M_2})\, .
\end{split}
\ee
We can now apply (\ref{ind1}) and set all $y_u = y$ and $z_u = z$, giving
\be
\label{eqarg}
\begin{split}
I_N(&\underbrace{y,\dots,y}_{M_1},\underbrace{z,\dots,z}_{M_2})
 =  \sum_{r=0}^\infty
\binom{M_2+r-1}{r}
\binom{M_1-r-\tfrac{1}{2}}{r+\tfrac{1}{2}}
\; y^{2r-1} \; z^{-2r} \\
&+ (-)^{M_1} \sum_{r=0}^\infty
\binom{M_1+r-1}{r}
\binom{N+r-\tfrac{1}{2}}{M_1+r+\tfrac{1}{2}}
\; y^{2(M_1+r)} \; z^{-2(M_1+r)-1}\, .\\
\end{split}
\ee
Now we take the limit $N \to \infty$ with $m_1=M_1/N$ and $m_2 = M_2/N$
fixed.
The terms in the second line are then negligible. Using the asymptotics
(\ref{ind1asym})
of the binomial coefficients, we obtain
\begin{equation}
\begin{split}
I_N(\underbrace{y,\dots,y}_{M_1},\underbrace{z,\dots,z}_{M_2})
&\xrightarrow{N\rightarrow\infty}
\frac{2}{\sqrt{\pi}}
N^{1/2}\,\frac{m_1^{1/2}}{y}
\sum_{r=0}^\infty
\binom{\tfrac{1}{2}}{r}
\left(\frac{m_2}{m_1}\right)^r \left(\frac{y}{z}\right)^{2r}\\
&=
\frac{2}{\sqrt{\pi}}
\left(\frac{M_1}{(\chi-a_1)^4} + \frac{M_2}{(\chi-a_2)^4}\right)^{1/2} \,
.
\label{asymres}
\end{split}
\end{equation}
The expression in the square root is the harmonic function $H$
appearing in the two-centre supergravity solution, where  the
instanton moduli $\chi^i$ are identified with the transverse coordinates
$y^i$ (up
to a dimensional scale). The large
$N$ limit of the function $I_N$ is thus identical, up to a numerical
factor,
to the classical supergravity volume element $\sqrt{\det g} = \sqrt H$.
The
above analysis straightforwardly generalises to a situation where the
vacuum
expectation values are degenerate at several centres.

Finally, we consider the behaviour of $I_N$ far away from the expectation
values,
$r\equiv|\chi| \gg \varphi_u$. Writing
\bea
x_u = (\chi-\varphi_u)^2 = r^2(1+\varepsilon_u) \,,\;\;\;
\varepsilon_u = - 2 \frac{\chi\cdot\varphi_u}{r^2} +
\frac{\varphi_u^2}{r^2}\,,
\eea
we have $\varepsilon \to 0$ in that region. From (\ref{indneg}) with
$d=-1$ we obtain
\bea
I_N(x_1,\dots,x_N) &=& \frac{1}{r^2}
\frac{S_\lambda(1+\varepsilon_u)}{
\left[\prod_u (1+\varepsilon_u) \right]
\left[\prod_{u<v} (2+\varepsilon_u+\varepsilon_v)\right]} \,.
\label{ineps}
\eea
The denominator can also be expressed in terms of Schur polynomials
\cite{macd},
\bea
\prod_u (1+\varepsilon_u) &=& \sum_{r=0}^n S_{(1^r)}(\varepsilon_u) \\
\prod_{u<v} (2+\varepsilon_u+\varepsilon_v) &=&
S_\delta(1+\varepsilon_u)\, ,
\eea
where the symbol $(1^r)$ denotes the partition (1, \dots, 1) of $r$, and $\delta=(N-1,
 N-2, \dots, 1)$. A formula by A.~Lascoux \cite{macd,lasc} provides an explicit Taylor
 expansion of arbitrary Schur polynomials around $x_u=1$,
\bea
\label{lascoux}
S_\lambda(1+\varepsilon_1,\dots,1+\varepsilon_N) &=&
\sum_{\mu\subset\lambda} d_{\lambda\mu}
S_\mu(\varepsilon_1,\dots,\varepsilon_N),
\eea
where the sum is over all partitions $\mu$ with $\mu_u \le \lambda_u$,
 all $u$, and the coefficients $d_{\lambda\mu}$ are given by
\bea
\label{dlm}
d_{\lambda\mu} &=&
\det \left[ \binom{\lambda_u+N-u}{\mu_v+N-v} \right]_{u,v=1,\dots,N}.
\eea
With the help of Mathematica, it is now possible to compute an expansion
of
$I_N(1+\varepsilon_u)$ in terms of Schur polynomials of the
$\varepsilon_u$,
\bea
I_N(1+\varepsilon_u) &=&
\underbrace{\frac{d_{\lambda 0}}{d_{\delta0}}}_{k_N}
\frac{1}{r^2}
\left(1+\frac{1}{N} S_{(1)}
+\frac{3N+1}{2N(N+1)}
\left(S_{(2)}-S_{(1,1)}\right)
+ \cdots \right)
\label{in1eps}
\eea
In order to compare with the supergravity calculation it turns out to be
more
useful to rewrite this expansion \cite{macd} in terms of power sums,
$p_k = p_k(\varepsilon) = \sum_{u=1}^N \varepsilon_u^k$,
\begin{equation}
\begin{split}
I_N(x_1,\dots,x_N)
&=
\frac{k_N}{r^2}
\left[
1-\frac{p_1}{N}
+\frac{(3N+2)p_2-p_1^2}{2N(N+1)}\right.\\
&
\left.
-\frac{(4N^2+9N+4)p_3-3(N+1)p_2p_1+p_1^3}{2N(N+1)(N+2)}
+O (\varepsilon^4)\right] \,.
\end{split}
\end{equation}
with the large-$N$ limit
\be
\label{inasym}
\begin{split}
I_N(x_1,\dots,x_N) \xrightarrow{N\to\infty}
\frac{1}{r^2} \frac{2 \sqrt{N}}{\sqrt\pi}
&\left[1-\frac{p_1}{N} +\frac{3Np_2-p_1^2}{2N^2}\right.\\
&\left.-\frac{4N^2p_3-3Np_2p_1+p_1^3}{2N^3}
+ O (\varepsilon^4)\right] \,.
\end{split}
\ee
On the supergravity side, the multi-centred harmonic
function with arbitrary locations $\varphi_u$ of the centres
has the following large-distance expansion,
\begin{equation}
\begin{split}
\sqrt{H} &= \left[ \sum_{u=1}^N\frac{1}{(\chi-\varphi_u)^4} \right]^{1/2}
= \frac{1}{r^2} \left[ \sum_{u=1}^N
  \frac{1}{(1+\varepsilon_u)^2} \right]^{1/2}\\
&= \frac{\sqrt{N}}{r^2} \left[
  1-\frac{p_1}{N}+\frac{3Np_2-p_1^2}{2N^2}
  -\frac{4N^2p_3-3Np_2p_1+p_1^3}{2N^3}
  + O (\varepsilon^4)
\right] \,.
\end{split}
\label{hlarger}
\end{equation}
This again coincides with the large-$N$ limit (\ref{inasym}) of
$I_N$.  We have
checked this equality up to eighth order in $\varepsilon$ so that
\bea
I_N(x_1,\dots,x_N) &\xrightarrow{N\to\infty}&
\frac{2}{\sqrt{\pi}} \sqrt{H}
+ r^{-2} O \left((\tfrac{\varphi_u}{r})^9\right)\, .
\label{equalas}
\eea

\end{document}